\documentclass[aps,prc,reprint,superscriptaddress,amsmath,amssymb,showpacs]{revtex4-1}
\usepackage[colorlinks=true,allcolors=blue]{hyperref}
\usepackage{graphicx}
\usepackage{physics}

\begin{document}
\title{Chiral magnetic currents with QGP medium response in heavy ion collisions at RHIC and LHC energies}

\author{Duan She}
\affiliation{College of Science, China Three Gorges University, Yichang 443002, China}
\author{Sheng-Qin Feng}
\email{email:fengsq@ctgu.edu.cn}
\affiliation{College of Science, China Three Gorges University, Yichang 443002, China}
\affiliation{Key Laboratory of Quark and Lepton Physics (Huazhong Normal Univ.), Ministry of Education,
Wuhan 430079, China}
\author{Yang Zhong}
\affiliation{College of Science, China Three Gorges University, Yichang 443002, China}
\author{Zhong-Bao Yin}
\affiliation{Institute of Particle Physics,Central China Normal University, Wuhan 430079, China}

\date{\today}

\begin{abstract}
We calculate the electromagnetic current with a more realistic approach in the RHIC and LHC
energy regions in the article. We take the partons formation time as the initial time of the magnetic field
response of QGP medium. The maximum electromagnetic current and the time-integrated current are two
important characteristics of the chiral magnetic effect (CME), which can characterize the intensity and
duration of fluctuations of CME. We consider the finite frequency response of CME to a time-varying
magnetic field, find a significant impact from QGP medium feedback, and estimate the generated electromagnetic
current as a function of time, beam energy and impact parameter.

\end{abstract}

\maketitle

\section{Introduction}\label{intro}
Quantum chromodynamics (QCD) is widely accepted to be the fundamental theory of the strong interactions. One remarkable feature of QCD is the existence of configurations of gauge fields characterized by different winding numbers. The chirality imbalance is induced by the nonzero topological charge through the axial anomaly\cite{Bell:1969ts,Adler:1969gk,Kharzeev:2009pj} of QCD as follows:
\begin{eqnarray}\label{eq:101}
\Delta N_{5}=\Delta(N_{R}-N_{L}) =-2N_{f}Q,
\end{eqnarray}
where $N_{f}$  is the number of flavors , and $\Delta N_{5}$  is the change in chirality ($N_{5}$ ) which is the difference between the number of modes with right- and left-handed chirality. In the limit of zero quark mass, $N_{5}$ is equal to the total number of particles plus antiparticles with right-handed helicity minus the total number of particles plus antiparticles with left-handed helicity. It has been proposed that if P- and CP-violating processes are taking place in the quark-gluon plasma (QGP) produced in relativistic heavy-ion collisions, then positive charges should separate from negative charges along the direction of angular momentum of the collision. The underlying mechanism is the so-called chiral magnetic effect (CME)\cite{Kharzeev:2009pj,Fukushima:2008xe,Kharzeev:2007jp}.
The CME is a particularly interesting phenomenon coming from the interplay of quantum  anomaly with magnetic field. The electric current introduced by the chirality imbalance \cite{Kharzeev:2009pj,Fukushima:2008xe,Kharzeev:2007jp} along an external magnetic field is as follows:
\begin{eqnarray}\label{eq:102}
\vec{J}=\sigma\vec{B}
\end{eqnarray}

\noindent where $\sigma=e^{2}\mu_{5}/(2\pi^{2})$  is the chiral magnetic conductivity with the chiral chemical potential $\mu_{5}$. One has to sum over the quark colors and flavors:
\begin{eqnarray}\label{eq:103}
\sigma=N_{c}\Sigma_{f}\frac{Q^{2}_{f}e^{2}\mu_{5}}{2\pi^{2}}
\end{eqnarray}

\noindent where $N_c$ is the color number of dynamical quarks. The CME leads to the event-by-event fluctuations of electric dipole moment of the QGP\cite{Kharzeev:2007tn,Kharzeev:2004ey,Burnier:2011bf}  in relativistic heavy ion collisions.

The question is whether CME exists in relativistic heavy ion collisions. Some analysis shows that the answer seems to be yes. Two necessary conditions for CME are chirality imbalance and strong magnetic field, which may be met in QGP generated in relativistic heavy ion collisions. Firstly, enormous magnetic field can be produced in non-central relativistic heavy-ion collision due to charged nucleus moving at speed close to the speed of light\cite{Mo:2013qya,Skokov:2009qp,Voronyuk:2011jd,Bzdak:2011yy,Tuchin:2013apa,Deng:2012pc}. Secondly, QCD which describes the behavior of the QGP allows topological charge changing transition that can induce chirality imbalance[5]. The magnetic field is the driving force and the electric charge separation is the manifestation of the CME\cite{Wang:2016mkm}. Therefore, the CME is very likely to exist in relativistic heavy - ion
collisions. Thus, heavy-ion collisions provide a unique terrestrial environment to investigate QCD matter in strong magnetic fields.

Over the past few years, the CME has been intensively explored by relativistic heavy ion collisions at the BNL Relativistic Heavy Ion Collider (RHIC) and at the CERN Large Hadron Collider (LHC), including STAR\cite{Adamczyk:2013kcb,Adamczyk:2014mzf,Wang:2012qs,PhysRevC.88.064911,Voloshin:2008jx,Abelev:2009ad,Abelev2009ac} , ALICE\cite{PhysRevLett.110.012301,Christakoglou:2011uqg} and for recent reviews see Refs.\cite{Kharzeev:2013ffa,Kharzeev:2015znc}. Observation of the chiral magnetic effect will be direct experimental evidence for the existence of topologically nontrivial gluon configurations.

One of the main issue to study the CME is the time evolution of the magnetic field in relativistic heavy ion collisions. This issue has been investigated by many works in the literature\cite{Kharzeev:2007jp,Mo:2013qya,Skokov:2009qp,Voronyuk:2011jd,Bzdak:2011yy,Tuchin:2013apa,Deng:2012pc}. The numerical computations executed by these works show that an enormous
magnetic field ($B\sim 10^{15}T$) can be generated at the very beginning of the collisions. However, according to these studies, the strength of the generated magnetic field decreases rapidly with time. If the lifetime of the magnetic field is so short that we can hardly see the imprint of the CME, and this is indeed a challenge for the manifestation of the CME in relativistic heavy-ion collisions. Nevertheless, It is argued that these studies of the CME are valid only at the early stage of the collision. It is argued \cite{Tuchin:2010vs,*Tuchin:2010vs2,McLerran:2013hla,Tuchin:2013ie,Zakharov:2014dia,Tuchin:2014iua,Tuchin:2015oka} that the magnetic field response from the QGP medium becomes increasingly important at a later time, and the magnetic field will maintain a much longer time in QGP.

In this paper, we precisely aim to address two pressing issues: (1) the time evolution of the strong magnetic field (which is the necessary driving force for CME), (2) the dynamical generation of CME current in response to time-dependent magnetic field. In order to make progress and to gain valuable insights on these problems, we chose the simplified model approach and  were able to obtain very interesting results.  We consider the finite frequency response of CME to a time-varying magnetic field, find significant impact from QGP medium feedback, and estimate the generated electromagnetic current as a function of time, beam energy and impact parameter, respectively.

This paper is organized as follows: the time evolution of the magnetic field in relativistic heavy-ion collisions is presented in Sec. II;  in Sec. III we discuss the chiral magnetic conductivity and the electromagnetic current; The results of the electromagnetic current in the RHIC and LHC energy regions are exhibited in Sec. IV. The conclusions are summarized in Sec. V.

\section{The magnetic field with the response of QGP medium} \label{QGPR}
In Ref.\cite{Kharzeev:2007jp}, Kharzeev, Mclerran, and Warringa (KMW) published an analytic model to calculate the magnetic field, and assumed a uniform nucleon density in nucleus rest frame in  relativistic heavy-ion collisions. In Ref. \cite{Mo:2013qya}, we improved the calculation of the magnetic field by using the Woods-Saxon nucleon distribution to replace uniform nucleon distribution. There also exist many other analyses using different methods\cite{Skokov:2009qp,Voronyuk:2011jd,Bzdak:2011yy,Tuchin:2013apa,Deng:2012pc} to calculate the magnetic field in the vacuum.

However, most of these analyses did not take into account the magnetic field response of the QGP medium which may obviously influence the time evolution of the magnetic field. Tuchin first analyzed\cite{Tuchin:2010vs,*Tuchin:2010vs2} the magnetic field feature in QGP medium, and concluded that the magnetic field was almost constant during the entire plasma lifetime due to high electric conductivity. Later, it was quantitatively studied in many works\cite{Tuchin:2010vs,*Tuchin:2010vs2,McLerran:2013hla,Tuchin:2013ie,Zakharov:2014dia,Tuchin:2014iua,Tuchin:2015oka} . To explore this problem, one needs to consider the electric conductivity $\sigma$  and chiral magnetic conductivity $\sigma_{\chi}$  which is induced by the CME. In Ref.\cite{McLerran:2013hla}, McLerran and Skokov found that the effects of finite $\sigma_{\chi}$  are not so important for the top RHIC and LHC energies. Therefore, we are not considering the effects of chiral magnetic conductivity in this paper. For electric conductivity $\sigma$, there are a lot of theoretical uncertainties\cite{Francis:2011bt,Ding:2010ga,Arnold:2003zc,Gupta:2003zh}.

We adopt the most optimistic assumption proposed in Ref.\cite{Deng:2012pc}, which assumes the electric conductivity $\sigma$  is large enough that one can take the QGP as an ideally conducting plasma. We use the following equations from Maxwell equations:
\begin{eqnarray}\label{eq:104-105}
\frac{\partial \vb{B}}{\partial t} &=& \nabla \times (\vb{v} \times \vb{B}),  \\
\vb{E} &=& - \vb{v} \times \vb{B},
\end{eqnarray}

\noindent where the $\vb{v}$ is the flow velocity of QGP.

To study how the magnetic field evolves in the QGP medium, we use the Bjorken picture for the longitudinal expansion\cite{Bjorken:1982qr,Ollitrault:2008zz} as:
\begin{equation}
v_z = \frac{z}{t},\label{eq:106}
\end{equation}

\noindent we adopt a linearized ideal hydrodynamic equation to describe the transverse velocities as
\begin{eqnarray}\label{eq:107-108}
v_x &=& \frac{c_s^2}{a_x^2} xt,  \\
v_y &=& \frac{c_s^2}{a_y^2} yt,
\end{eqnarray}

\noindent where $a_{x,y}$ are the root-mean-square widths of the transverse distribution and $c_s$ is the speed of sound. We take $c_s^2 \sim 1/3$ and $a_x \sim a_y \sim 3$,respectively.

We can calculate the magnetic field $\vb{B}(t)$ at a given initial condition $\vb{B}^{0}(\vb{r})=\vb{B}(t=t_{0},\vb{r})$ by substituting the velocity into  Eqs.(4)-(5),
where $t_{0}$ is the formation time of partons. It is found that only the $y$ component of the magnetic field of the central point of two nuclei collision remains.
Therefore, we only consider the $y$ component of the magnetic field at the
center of the collision region $\vb{r}=\vb{0}$, and one gets the following solution
\begin{equation} \label{eq:109}
B_y(t\geq t_{0},\vb{0}) = \frac{t_0}{t} e^{-\frac{c_s^2}{2a_x^2}(t^2 - t_0^2)} B_y^0(\vb{0}).
\end{equation}

We will calculate the magnetic field $B_{y}(\vb{0})$ at $t\leq t_{0}$ at the central point ($\vec{x}=0$) in which case
it is pointing in the $y$ direction, and it is given by contributions of participant and spectator nucleons in relativistic heavy-ion collisions.
\begin{equation} \label{eq:110}
B_{y}(t\leq t_{0},\vb{0}) =B_{ys}^{+}(t,\vb{0})+B_{ys}^{-}(t,\vb{0})+B_{yp}^{+}(t,\vb{0})+B_{yp}^{-}(t,\vb{0}).
\end{equation}

\noindent where $B_{ys}^{\pm}(t,\vb{0})$ and $B_{yp}^{\pm}(t,\vb{0})$  are the contributions of the spectators and
the participants moving in the positive or negative $z$ direction,
respectively. For spectators, we assume that they do not scatter
at all and that they keep traveling with the beam rapidity $Y_0$. The magnetic field  from spectators is given as:
\begin{eqnarray}\label{eq:111}
B_{ys}^{\pm}(t,\vb{0})&&=\pm Z\alpha_{EM}\sinh {Y_0}
\int_{V_{s}^{\pm}}{\rm d}^3\vec{x}^\prime\rho_{\pm}(\vec{x}^\prime)\nonumber\\
&&\times\frac{\vec{x}^\prime_\bot\times\vec{e}_{z'}}
{[\vec{x}^{\prime 2}_\bot+(\pm t \sinh Y_0+ z'\cosh Y_0)^{2}]^{3/2}},
\end{eqnarray}

\noindent where $\rho_{\pm}(\vec{x}^\prime)$ is the three dimension Wood-Saxon nuclear density:
\begin{eqnarray}
\rho_{\pm}(\vec{x}^\prime)=
\frac{\gamma n_0}{1+\exp\left(\frac{\sqrt{(x'\mp{b/2})^2+y'^{2}+(\gamma z')^{2}}-{\rm R}}{d}\right)}£¬
\label{eq:112}
\end{eqnarray}

\noindent where $\gamma$ is the Lorentz factor, $n_{0}=0.17 \mathrm{fm}^{-3}$, $d = 0.54 \mathrm{fm}$, and the radius $R=1.12A^{1/3} \mathrm{fm}$. The contribution of the participants to the magnetic field can be also given by
\begin{eqnarray}
B_{yp}^{\pm}(t,\vb{0}) &&=\pm Z\alpha_{EM}
\int_{V_{p}^{\pm}}{\rm d}^3\vec{x}^\prime \int_{-Y_0}^{Y_0}{\rm d}Y f(Y){\sinh Y}\rho_{\pm}(\vec{x}^\prime)\nonumber\\
&&\times\frac{\vec{x}^\prime_\bot\times\vec{e}_{z'}}
{[\vec{x}^{\prime 2}_\bot+(\pm t \sinh Y+ z^\prime\cosh Y)^{2}]^{3/2}},
\label{eq:113}
\end{eqnarray}

\noindent where
\begin{eqnarray}
f(Y)=\frac{a}{2\sinh(aY_0)}{\rm e}^{aY},  \hskip1cm -Y_{0}\leq{Y}\leq{Y_{0}},
\label{eq:114}
\end{eqnarray}
\noindent and experimental data show that $a\approx1/2$, consistent with the baryon junction stopping mechanism.

To study the time evolution of magnetic field from Eq.(9), we must know the formation time $t_0$  of partons and the initial magnetic field $B_y^0(\vb{0})$ at this time. For the initial magnetic field $B_y^0(\vb{0})$, we can
calculate from Eq.(10) to Eq.(14) which developed from  the method in Ref. \cite{Mo:2013qya}. For the formation time $t_0$ of partons~\cite{Kharzeev:2000ph}, the following approximation formula has been used:
\begin{equation}
t_0 \simeq 1 / Q_{s},
\label{eq:115}
\end{equation}
\noindent where $Q_s$ is the saturation momentum. According to saturation analysis in Refs.\cite{Kharzeev:2000ph,Kowalski:2007rw,Lappi:2009mp}, $Q_s$ is related to collision nuclei and collision energy as:
\begin{equation} \label{eq:116}
Q_{s}^{2} \sim A^{1/3}x^{-\lambda}, x=Q_{s}/\sqrt{s},
\end{equation}

\noindent where A is the atomic quantity of collision nuclei, and $\lambda$ is between $0.25$ and $0.3$, we give $\lambda=0.28$ in the paper. Then we get
\begin{equation} \label{eq:117}
Q_{s}^{2} =k A^{(\frac{2}{3(2+\lambda)})}\sqrt{s}^{(\frac{2\lambda}{2+\lambda})},
\end{equation}

\noindent where $k$ is coefficient of proportionality, we take it as a constant. These $Q_{s}^{2}(\sqrt{s} =130\,\mathrm{GeV}, b, A=197)$ were calculated by Tab.2 in Refs.\cite{Kharzeev:2000ph}
for different impact parameters at RHIC $\sqrt{s} =130\,\mathrm{GeV}$ of Au-Au collisions. By comparing with the results from $\sqrt{s} =130\,\mathrm{GeV}$ of Au-Au collisions, we
compute $Q_{s}^{2}(\sqrt{s},b,A)$ at different $b$, different collision energy $\sqrt{s}$ and different collision nuclei $A$ as:

\begin{eqnarray} \label{eq:118}
Q_{s}^{2}(\sqrt{s},b,A) &=& (\frac{A}{197})^{\frac{2}{3(2+\lambda)}}(\frac{\sqrt{s}}{130})^{\frac{2\lambda}{2+\lambda}} \\*
&\times& Q_{s}^{2}(\sqrt{s}=130 GeV,b,A = 197).\nonumber
\end{eqnarray}

From Eq.(15) to Eq.(18), we can calculate the formation time as $t_{0}=0.16\,\mathrm{fm}$  for $\sqrt{s}=200\,\mathrm{GeV}$ Au-Au collisions with $b = 8\,\mathrm{fm}$, $t_{0}=0.11\,\mathrm{fm}$  for $\sqrt{s}=2760\,\mathrm{GeV}$  Pb-Pb collisions and $t_{0}=0.10\,\mathrm{fm}$  for $\sqrt{s}=5020\,\mathrm{GeV}$  Pb-Pb collisions with $b = 8\,\mathrm{fm}$. Figure 1(a) provides the comparison of the time evolution of the magnetic field at the central point with or without considering the QGP response for Au-Au collisions with $b = 8\,\mathrm{fm}$ at $\sqrt{s}=200\,\mathrm{GeV}$. fig. 1(b) and fig. 1(c)  are same as the fig.1(a) but for Pb-Pb collisions at $\sqrt{s} = 2760\,\mathrm{GeV}$ and for Pb-Pb collisions at $\sqrt{s} = 5020\,\mathrm{GeV}$, respectively. It is found that the magnetic field considering the QGP response will maintain a much longer time than that in the vacuum.

\begin{figure}
\includegraphics[width=8.5cm]{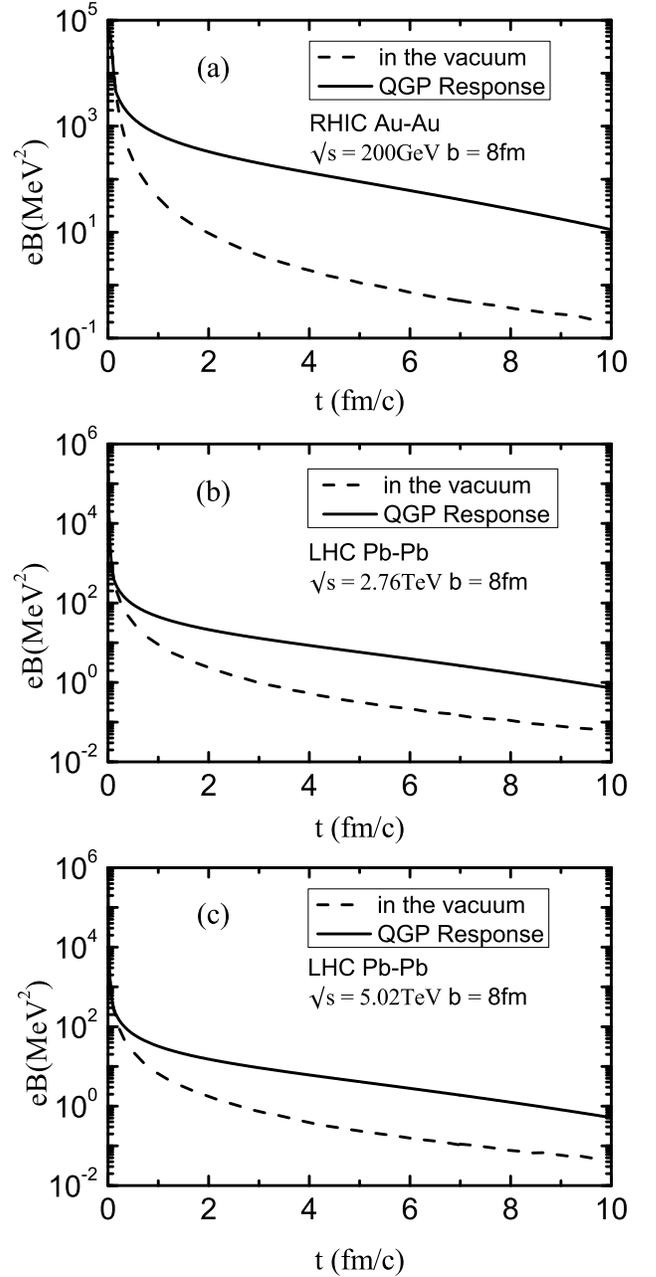}
\caption{\label{fig1}The time evolutions of magnetic field at central point for $\sqrt{s}=200\,\mathrm{GeV}$ Au-Au collisions with $b = 8\,\mathrm{fm}$, $\sqrt{s} = 2760\,\mathrm{GeV}$ Pb-Pb collisions with $b = 8\,\mathrm{fm}$ and $\sqrt{s} = 5020\,\mathrm{GeV}$ Pb-Pb collisions with $b = 8\,\mathrm{fm}$. The solid line and dashed line represent with and without considering the response of QGP medium, respectively.}
\end{figure}

\begin{figure}
\includegraphics[width=8.5cm]{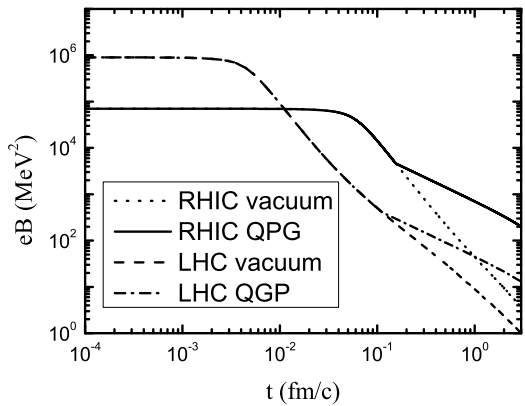}
\caption{\label{fig2}The comparisons of magnetic field time evolution  at central point between for Au-Au collisions with $b = 8\,\mathrm{fm}$ at $\sqrt{s}=200\,\mathrm{GeV}$ and for
Pb-Pb collisions with $b = 8\,\mathrm{fm}$ at $\sqrt{s} = 2760\,\mathrm{GeV}$ with and without considering QGP response.}
\end{figure}

Figure 2 shows a comparison of magnetic field time evolution at central point between for Au-Au collisions with $b = 8\,\mathrm{fm}$ at $\sqrt{s}=200\,\mathrm{GeV}$ and for
Pb-Pb collisions with $b = 8\,\mathrm{fm}$ at $\sqrt{s} = 2760\,\mathrm{GeV}$ with and without considering QGP response. It is found that that the magnetic field  $B_{t=0}$ at LHC $\sqrt{s} = 2760\,\mathrm{GeV}$
is about 100 times that at RHIC $\sqrt{s}=200\,\mathrm{GeV}$, but the magnetic field at LHC $\sqrt{s} = 2760\,\mathrm{GeV}$ decreases much faster than that at RHIC $\sqrt{s}=200\,\mathrm{GeV}$. The formation time $t_{0}$ varies little  with the collision energy $\sqrt{s}$. On the whole, the magnetic field at RHIC is far larger than the magnetic field at LHC in the case of $t>t_{0}$.

\begin{figure}
\includegraphics[width=8.5cm]{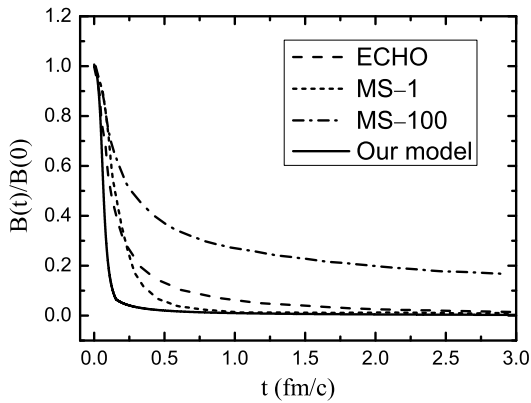}
\caption{\label{fig3}The comparison of various results for the time dependence of the magnetic field normalized by the peak value of the magnetic field from
different studies: ECHO-QGP\cite{Inghirami:2016iru} (dashed line), McLerran-Skokov\cite{McLerran:2013hla} with different electric conductivity (dotted line
curve for $\sigma=\sigma_{LQCD}$, dash-dotted line for $\sigma=100\sigma_{LQCD}$).}
\end{figure}

One potentially very useful application of the results on time-dependent magnetic fields (in Fig.1) would be serving as an input for recently developed anomalous hydrodynamic simulations for the CME\cite{Jiang:2016wve,Shi:2017cpu}. As shown in Fig.2 of Ref.\cite{Shi:2017cpu}, the final CME signal very sensitively depends on the time-dependence of magnetic field.
Many calculations have been done to study the strength and space-time dependence of the magnetic field which are among the most crucial factors in quantifying the CME signal. However, the time evolution of magnetic field (B) remains an open question. The hot medium created in the collision could delay the decrease of the magnetic field through the generation of an induction current in response to the rapidly decaying magnetic field. In fig.3, we show a comparison of various results for the time dependence of the magnetic field normalized by the peak value of the magnetic field: the study by McLerran-Skokov\cite{McLerran:2013hla} with conductivity  $\sigma=\sigma_{LQCD}$ and $\sigma=100\sigma_{LQCD}$, ECHO-QGP simulation\cite{Inghirami:2016iru} and our model. Compared with the other results, the magnetic field given by us is the fastest to decrease with time.

\section{CHIRAL MAGNETIC CONDUCTIVITY AND CHIRAL MAGNETIC CURRENT} \label{QGPR}
Let us now study the induced current by a magnetic field created in relativistic heavy ion collisions. As a qualitative research purpose, one assumes that the magnetic field
is homogeneous in space. The induced current can be given as follows\cite{Kharzeev:2009pj}:
\begin{equation} \label{eq:119}
j(t)=\int_{0}^{\infty}\frac{dw}{\pi}[\sigma'_{\chi}(\omega)\cos(\omega t)+\sigma''_{\chi}(\omega)\sin(\omega t)]\tilde{B}(\omega),
\end{equation}

\noindent where
\begin{eqnarray}\label{eq:120}
\tilde{B}(\omega)=\int_{t_{0}}^{\infty}dte^{i\omega t}B(t),
\end{eqnarray}

\noindent and $\sigma'_{\chi}(\omega)$, $\sigma''_{\chi}(\omega)$ are the real part and imaginary part of the chiral magnetic conductivity. The so-called Kramers-Kroning relation gives the relationship between
the real and imaginary parts as follows:
\begin{eqnarray}\label{eq:121-122}
\sigma'_{\chi}   (\omega) = \frac{1}{\pi}  \textbf{P} \int_{-\infty}^{\infty} dq_{0} \frac{\sigma''_{\chi}(q_{0})}{q_{0}-\omega}, \\
\sigma''_{\chi}(\omega)=-\frac{1}{\pi}\textbf{P}\int_{-\infty}^{\infty}dq_{0}\frac{\sigma'_{\chi}(q_{0})}{q_{0}-\omega},
\end{eqnarray}
where $\sigma_{\chi}(\omega)=\lim_{\vec{p}\rightarrow0}\sigma_{\chi}(p_{0}=\omega,\vec{p})$ . The symbol $\textbf{P}$ of Eqs.(21) and (22) denote the
integral which avoids the singularity via the upper and the lower complex plane. The chiral magnetic conductivity will be complex as follows:
\begin{equation}  \label{eq:123}
\sigma_{\chi}(p)=\sigma'_{\chi}(p)+i\sigma''_{\chi}(p)
\end{equation}

\noindent where $\sigma'_{\chi}(p)$  and $\sigma''_{\chi}(p)$ are real functions. They can be expressed as:
\begin{eqnarray} \label{eq:124-125}
\sigma'_{\chi}(p)=\frac{1}{p^{i}}\Im G^{i}_{R}(p), \\
\sigma''_{\chi}(\omega)=-\frac{1}{p^{i}}\Re G^{i}_{R}(p),
\end{eqnarray}

\noindent where $G^{i}_{R}(p)=\frac{1}{2}\varepsilon^{ijk}\tilde{\prod}_{R}^{jk}(p)$ is the retarded correlator.

The retarded correlator $G^{i}_{R}(p)$  is a very important physical quantity in linear response theory of chiral magnetic conductivity. The detailed analysis and calculation process of $G^{i}_{R}(p)$
can be found in Ref.\cite{Kharzeev:2009pj}. The retarded correlator can be given as follows:
\begin{eqnarray} \label{eq:126}
G^{i}_{R}(p) &=& \frac{ie^{2}}{16\pi^{2}}\frac{p^{i}}{p} \frac{p^{2}-p_{0}^{2}}{p^{2}}\int_{0}^{\infty} \dd{q}[f(q)  \\*
& \times & \sum_{t=\pm}(2q+tp_{0})\log\frac{(p_{0}+i\varepsilon+tq)^{2}-(q+p)^{2}}{(p_{0}+i\varepsilon+tq)^{2}-(q-p)^{2}}], \nonumber
\end{eqnarray}

\noindent where
\begin{equation}\label{eq:127}
f(q)= \sum_{s=\pm}s[\tilde{n}(q-\mu_{s})-\tilde{n}(q+\mu_{s})],
\end{equation}

\noindent where $\tilde{n}(x)=[1+\exp(\beta x)]^{-1}$ is the Fermi-Dirac distribution function. To derive the imaginary part of the chiral magnetic conductivity,
one can express the imaginary part of the logarithm in Eq.(21) with $q\geq0$ and $p=|\vec{p}|\geq0$  as follows:
\begin{eqnarray} \label{eq:128}
&\Im &\sum_{t=\pm}(2q+tp_{0})\log\frac{(p_{0}+i\varepsilon+tq)^{2}-(q+p)^{2}}{(p_{0}+i\varepsilon+tq)^{2}-(q-p)^{2}} \\*
&& \quad =\pi[2q-|p_{0}|\theta(p_{0}^{2}-p^{2})][\theta(q_{+}-q)-\theta(q_{-}-q)] \nonumber \\
&& \quad +\pi p_{0}\theta(p^{2}-p_{0}^{2})[\theta(q-q_{+})-\theta(q-q_{-})], \nonumber
\end{eqnarray}

\noindent where $q_{\pm}=\frac{1}{2}|p_{0}\pm p|$.

The real and imaginary parts of chiral magnetic conductivity can be computed by using the Kramers-Kroning relation as shown in Eqs.(21) and (22). Fig.4(a, b) shows the full frequency and momentum
dependence of $\sigma_{\chi}(\omega,p)$. We can find some features of the magnetic conductivity as follows:

(1)	There is a peak at $p=0$  and $\omega \approx 5.406T$  at high temperature, which tends to disappear when $p>0$.

(2)	 The conductivities are not vanishing at $p=0$ and $\omega\neq 0$, and they still present a discontinuity at $p=0$ and $\omega=0$.

(3)	 The real part of conductivities can be negative note that the phase angle is between $0$ and $\pi$. When $p=\omega$, the imaginary part of conductivities is equal to zero.

(4)	 The chiral magnetic conductivity is approximately vanishing at high temperature, in the regime $\omega\gg p$ or $p\gg\omega$.

\begin{figure}
\includegraphics[width=8.5cm]{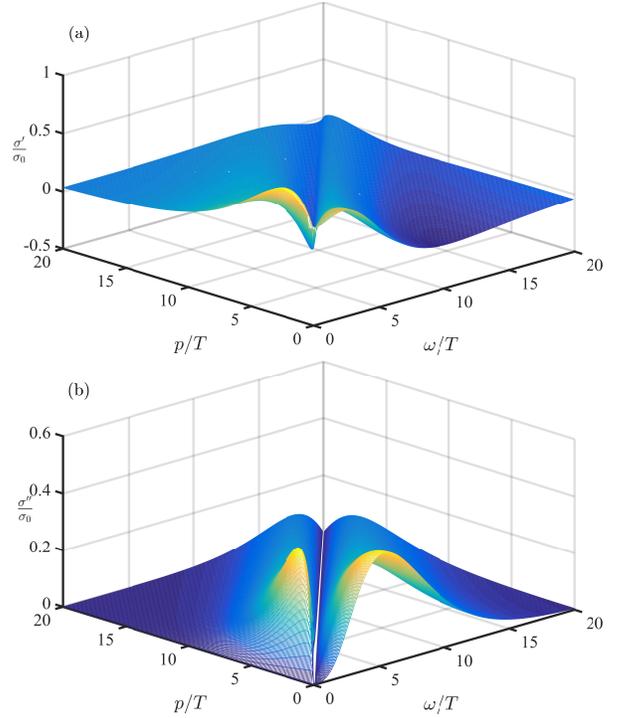}
\caption{\label{fig4}Frequency and momentum dependence of the chiral magnetic conductivity normalized to the zero frequency-momentum value, from a numerical
evaluation. (a) shows the real part of chiral magnetic conductivity, and (b) shows the imaginary part of the conductivity. Where $\mu=10 \mathrm{MeV}$,
$\mu_{s}=1 \mathrm{MeV}$ , and the temperature $T=200 \mathrm{MeV}$.}
\end{figure}

\section{THE RESULTS OF THE ELECTROMAGNETIC CURRENT} \label{QGPR}
In the above chapter, we have computed the real and imaginary parts of the magnetic conductivity, which are shown in fig.4. And then we will use the Eq.(19) to calculate the electromagnetic current. It can be
seen from the Eqs.(19-20)  that in order to calculate  the electromagnetic current we must study the dependence of the magnetic field  on time after the formation of the parton. The detailed evolution formula of the magnetic field with time after $t\geq t_{0}$ is given by Eq.(9). The dependence of the magnetic field on time before the formation of the part ($t\leq t_{0}$)is given by Eqs. (10-14).

The dependencies of the electromagnetic current on time with or without the QGP response in the RHIC and LHC are shown in Fig.5. The time dependencies of chiral magnetic currents and magnetic field are calculated by using Eqs.(19-23) and Eq.(9).  It is found that the strength of the electromagnetic current considering the QGP response is larger than that of the electromagnetic current in the vacuum. In order to make a related study of the experimental results of CMS collaboration in Refs.\cite{Sirunyan:2017quh}, we also compute the electromagnetic current at LHC $\sqrt{s}= 5020 \mathrm{GeV}$ for Pb - Pb collisions with $b=8 \mathrm{fm}$.
\begin{figure}
\includegraphics[width=8.5cm]{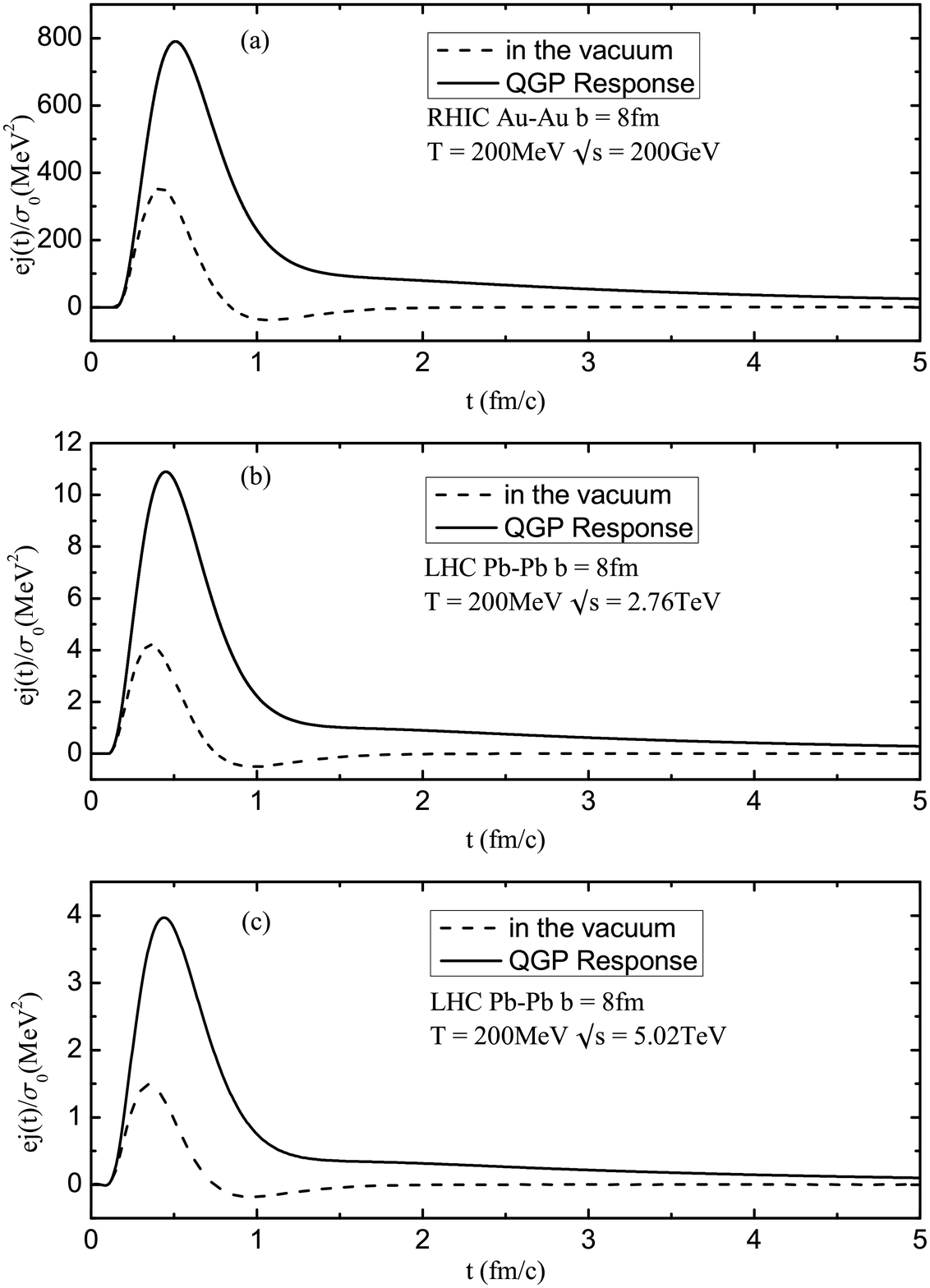}
\caption{\label{fig5}The dependencies of the electromagnetic current on the time with the QGP response or in the vacuum in the RHIC $\sqrt{s}= 200 \mathrm{GeV}$
and in the LHC $\sqrt{s}= 2760 \mathrm{GeV}$ and $\sqrt{s}= 5020 \mathrm{GeV}$,
the solid line represents the result of considering the QGP response, and the dashed line represents the result in the vacuum.}
\end{figure}

The comparisons of the dependencies of the electromagnetic current on the time in the RHIC $\sqrt{s}= 200 \mathrm{GeV}$,  and in the LHC $\sqrt{s}= 2760 \mathrm{GeV}$ and $\sqrt{s}= 5020 \mathrm{GeV}$ are given in Fig.6(a,b).
It is found that the strength of the electromagnetic current at RHIC $\sqrt{s}= 200 \mathrm{GeV}$ for Au-Au collisions with $b=8 \mathrm{fm}$  far outweigh that of the electromagnetic current at  LHC $\sqrt{s}= 2760 \mathrm{GeV}$ and $\sqrt{s}= 5020 \mathrm{GeV}$ for Pb - Pb collisions with $b=8 \mathrm{fm}$. The maximum
electromagnetic current  can reach $800 \mathrm{MeV}^{2}$ at RHIC $\sqrt{s}= 200 \mathrm{GeV}$, but the maximum
electromagnetic current  is only $11 \mathrm{MeV}^{2}$ at LHC $\sqrt{s}= 2760 \mathrm{GeV}$ and then decreases
to $4 \mathrm{MeV}^{2}$ at LHC $\sqrt{s}= 5020 \mathrm{GeV}$. From Fig.(5) and Fig.6, it is found that the electromagnetic current in the LHC energy region is so small that it is difficult to study CME. The main reason for such a small electromagnetic current in the LHC energy region is due to the sharply decrease of the magnetic field with time in the LHC energy region. Such observation is important for understanding why the recent CMS measurements \cite{Sirunyan:2017quh} at LHC $\sqrt{s}= 5020 \mathrm{GeV}$ see no CME signal.

The study of the CME should be mainly concentrated in the RHIC energy region, so we have carried out the study of the dependencies of electromagnetic current on collision energy, impact parameter and temperature in the RHIC energy region.
\begin{figure}
\includegraphics[width=8.5cm]{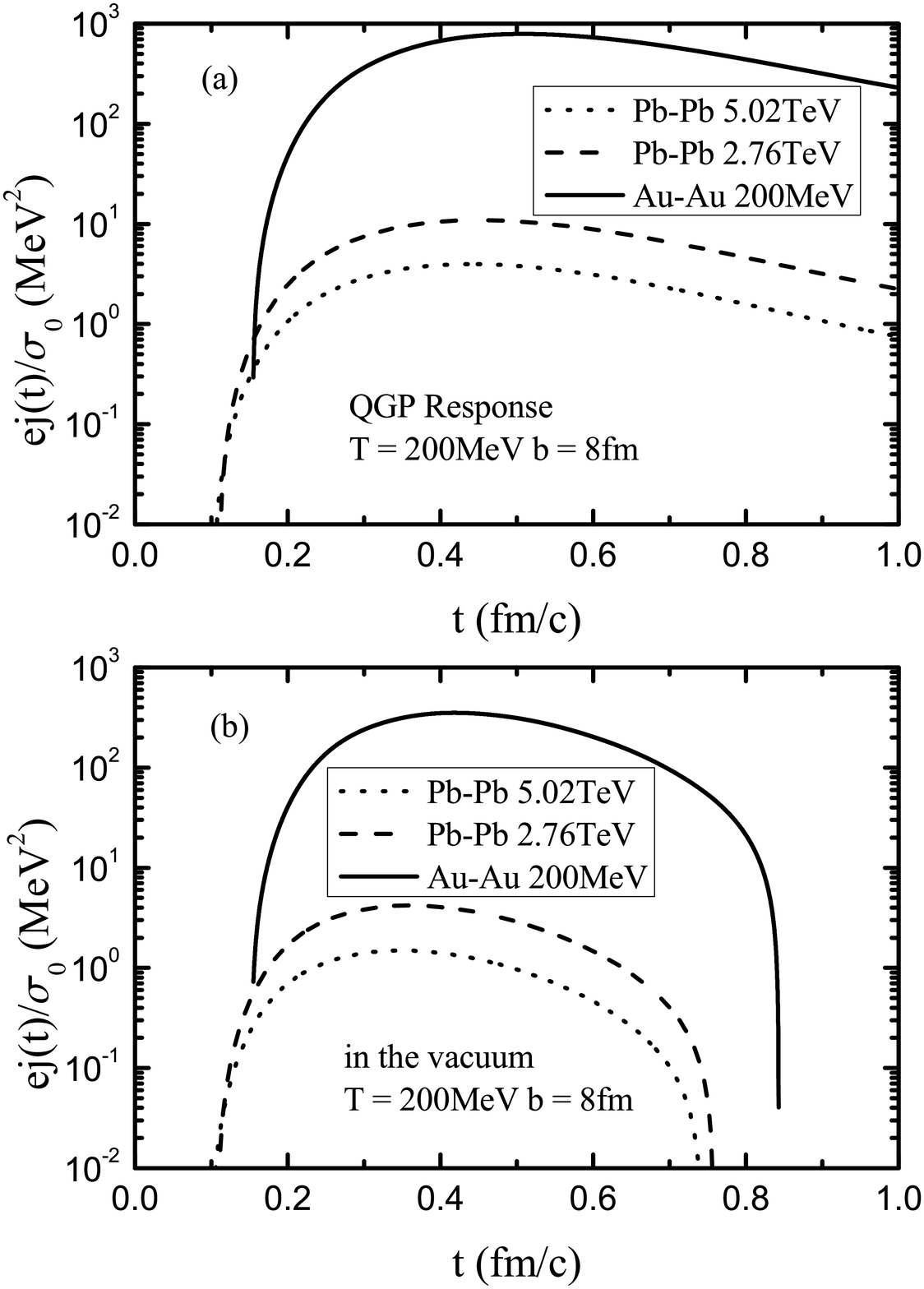}
\caption{\label{fig7}The comparisons of the dependencies of the electromagnetic current on the time with the QGP response (Fig.7(a)) or in the vacuum (Fig.7(b))in the RHIC $\sqrt{s}= 200 \mathrm{GeV}$,
 LHC $\sqrt{s}= 2760 \mathrm{GeV}$ and $\sqrt{s}= 5020 \mathrm{GeV}$. The solid line represents the result at RHIC $\sqrt{s}= 200 \mathrm{GeV}$ for Au-Au collisions, the dashed line
 represents the result at LHC $\sqrt{s}= 2760 \mathrm{GeV}$ and the dotted line represents the result at LHC $\sqrt{s}= 5020 \mathrm{GeV}$.}
\end{figure}

It is an interesting subject to study the relationship between electromagnetic current and magnetic field. It is well known that the magnetic field decreases monotonically with time in relativistic heavy ion collisions. A magnetic field in relativistic heavy ion collisions is a uniform function of time.  As can be seen from Fig.7, the electromagnetic current increases with time at first, and then decreases with the increase of time. By comparing the dependence of the electromagnetic current and the magnetic field on the time variation, respectively, we can establish the relationship between the electromagnetic current and the magnetic field.
\begin{figure}
\includegraphics[width=8.5cm]{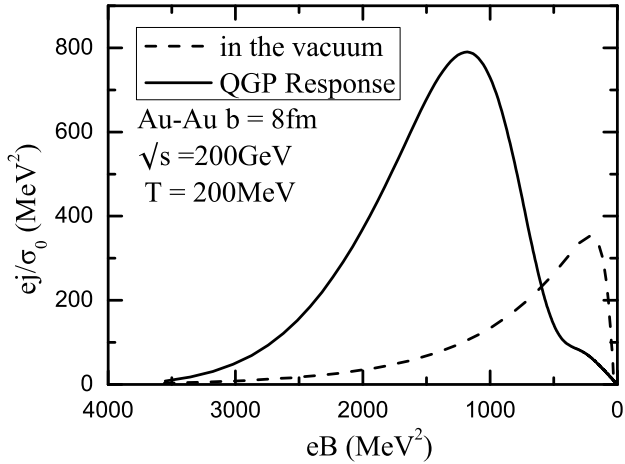}
\caption{\label{fig7}The dependence of the electromagnetic current on the magnetic field at RHIC $\sqrt{s}= 200 \mathrm{GeV}$ for Au-Au collisions with $b=8 \mathrm{fm}$. The solid line represents the result by considering the QGP response, and the dashed line represents the result in the vacuum. The formation time is given as $t_{0}=0.16 \mathrm{fm}$.}
\end{figure}

Fig.7 shows the dependence of the electromagnetic current on the magnetic field  by considering the QGP response and in the vacuum at RHIC $\sqrt{s}= 200 \mathrm{GeV}$ for Au-Au collisions with $b = 8 \mathrm{fm}$.
It is found that the fast decaying magnetic field can give rise to a non-negligible current which firstly increases to the maximum at $eB\sim 1.25\times 10^{3} \mathrm{MeV^{2}}$($eB\sim 2.51\times 10^{2} \mathrm{MeV^{2}}$ ) considering QGP response (in the vacuum), then diminishes. One also finds that the electromagnetic current considering the QGP response is larger than the electromagnetic current in the vacuum in most of the magnetic field regions.

\begin{figure}
\includegraphics[width=8.5cm]{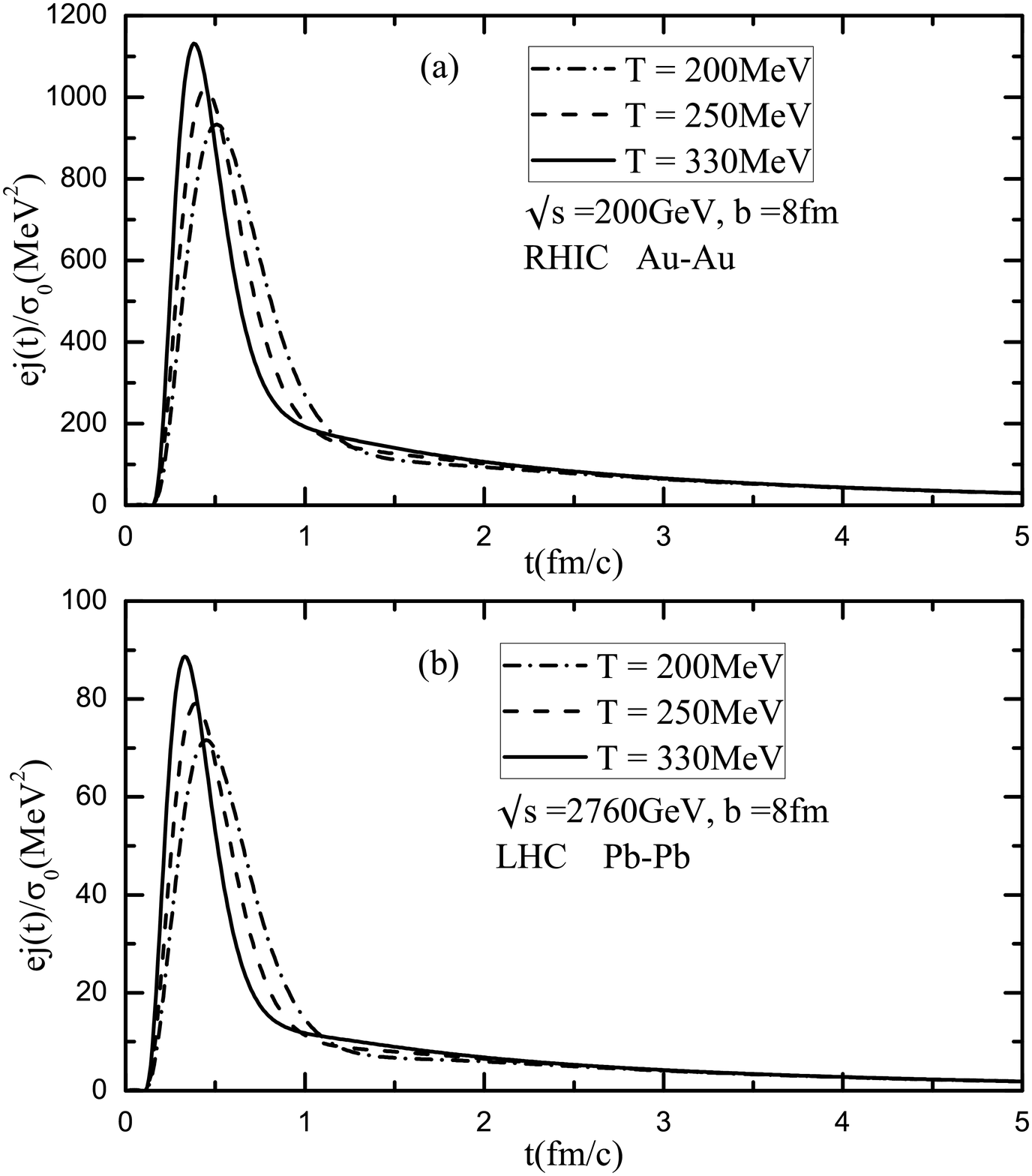}
\caption{\label{fig8}Time dependence of the induced electromagnetic current normalized to zero frequency chiral magnetic conductivity. The results are plotted with different values of the temperature above the QCD phase transition. Top (a): Au-Au collisions with $\sqrt{s}= 200 \mathrm{GeV}$  and $b=8 \mathrm{fm}$  in RHIC energy region. Bottom (b): Pb-Pb collisions with $\sqrt{s}= 2760 \mathrm{GeV}$ and $b=8 \mathrm{fm}$ in LHC energy region. }
\end{figure}

In fig.8 it can be seen that even in the non-interacting case there still is sizable response at high temperature. Hence even for such fast changing fields there will really be an induced current. One finds that thermal fluctuations will increase in magnitude when the temperature is increased. These fluctuations can cause the spins of the particles to align along the fast decaying magnetic field, it clearly takes some time for the current to respond. By keeping $\sqrt{s}, b$ fixed and increasing the temperature, the system will respond faster to the changing magnetic field and the maximal current will be larger.

 We denote the $t_{\textrm{response}}$  as the response time of the electromagnetic current from the beginning of the current production to the maximum value $j_{\textrm{max}}$ , and  as the maximum electromagnetic current of the system. The maximum electromagnetic current  and the response time  are important physical quantities, and they represent the intensity of the electromagnetic signal and the speed of the response of the CME of the system.

As mentioned above, CME is mainly in the RHIC energy region, so below we will make a detailed analysis of the characteristics of the electromagnetic current in the RHIC energy region. We will carry out the study of the dependencies of electromagnetic current on collision energy, impact parameter and temperature in the RHIC energy region. In order to study the dependencies of electromagnetic current on collision energy ($\sqrt{s}$), we calculate the energy dependence of formation time $t_{0}$  and magnetic field $eB^{0}_{y}(0)$ at $t=t_{0}$ at central point
$\vb{r}=\vb{0}$ for Au-Au collisions as shown in table 1.

\begin{table}
\caption{\label{tab:CuRHICdata}Energy dependence of formation time $t_{0}$  and initial magnetic field $eB^{0}_{y}(0)$ at $t=t_{0}$  for $b=8 \mathrm{fm}$ Au-Au collisions }
\begin{ruledtabular}
\begin{tabular}{lll}
 $\sqrt{s} (\mathrm{GeV}) $  & $t_0 (\mathrm{fm})$ & $eB_y^0 (\mathrm{MeV}^2) $ \\
 \hline
  7.7  &  0.231  &  $1.03 \times 10^{4}$ \\
  10 & 0.224  & $1.30 \times 10^{4}$  \\
 20 &  0.206 &  $2.01 \times 10^{4}$  \\
 30 & 0.196 & $2.19 \times 10^{4}$  \\
  40 & 0.189 & $2.09 \times 10^{4}$ \\
  50 &  0.184 & $1.89 \times 10^{4}$  \\
60 & 0.180 & $1.67 \times 10^{4}$ \\
70 &  0.176 & $1.47 \times 10^{4}$  \\
80 & 0.174 & $1.29 \times 10^{4}$ \\
90 &  0.171  & $1.13 \times 10^{4}$  \\
100 & 0.169 &  $1.00 \times 10^{4}$  \\
110 & 0.167 & $8.65 \times 10^{3}$ \\
120 & 0.165 & $7.98 \times 10^{3}$ \\
130 & 0.164 & $7.18 \times 10^{3}$ \\
140 & 0.162 & $6.50 \times 10^{3}$ \\
150 & 0.161 & $5.90 \times 10^{3}$ \\
160 & 0.160 & $5.39 \times 10^{3}$  \\
170 & 0.158 & $4.94 \times 10^{3}$ \\
180 &  0.157 & $4.55 \times 10^{3}$  \\
190 & 0.156 & $4.20 \times 10^{3}$  \\
200 & 0.155 & $3.89 \times 10^{3}$  \\
\end{tabular}
\end{ruledtabular}
\end{table}

The dependence of the maximum electromagnetic current on the center-of-mass energy is presented in fig. 9. Figure 9 shows that it is not a monotonic change of the dependence of the maximum electromagnetic current
on the center-of-mass energy. Closely related to the beam energy dependence, the results of fig.9 also demonstrate a strong decrease with increasing collision energy as $\sqrt{s} > 30 \mathrm{GeV}$. Figure 9 also suggests that the CME signal could vanish at LHC energies.

The dependence of the time-integrated current signal ($Q = \int j(t)dt$) on the center-of-mass energy is shown in fig.10. We find that the time-integrated current signal reaches the maximum at $\sqrt{s}\approx 30 \mathrm{GeV}$, then decreases with the increase of $\sqrt{s}$. The relation of the time-integrated current with the energy change is consistent with the maximum electromagnetic current with the collision energy.

From fig. 10, we can find that the peak value of time-integrated current appears near $\sqrt{s}\approx 30 \mathrm{GeV}$, and then decreases with the increase of energy, which indicates that the CME at this collision energy $\sqrt{s}\approx 30 \mathrm{GeV}$ is the most obvious. The qualitative trend of fig.9 and fig.10 is in agreement with STAR BES analysis results for a wide range of beam energy\cite{Adamczyk:2014mzf}, which are based on the two-component decomposition method given by Refs.\cite{Bzdak:2012ia,Liao:2014ava}.

\begin{figure}
\includegraphics[width=8.5cm]{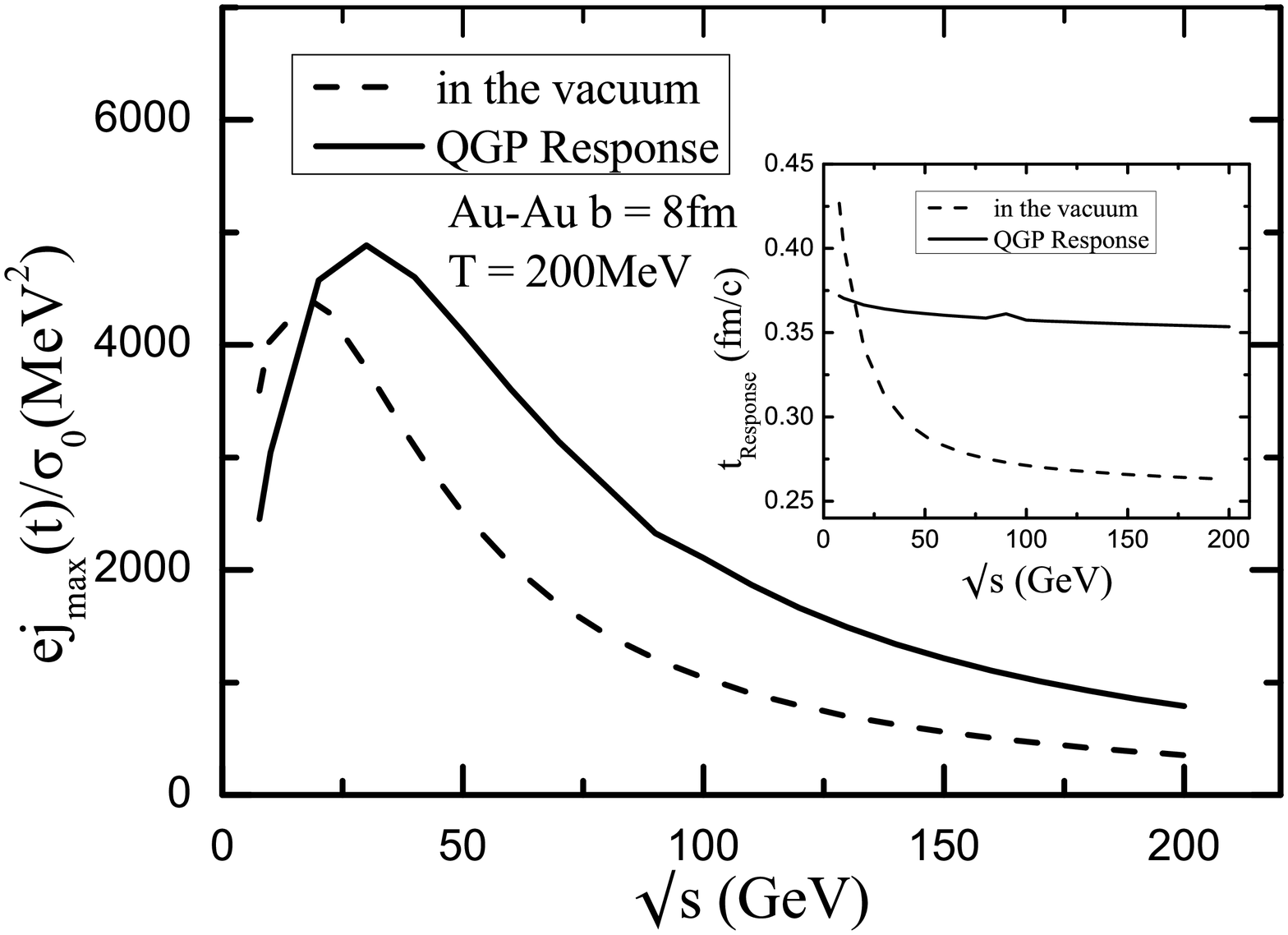}
\caption{\label{fig9}The dependence of the maximum electromagnetic current on the central-of-mass energy for  Au-Au collisions with $T = 200 \mathrm{MeV}$  and $b=8 \mathrm{fm}$, the solid line represents the  result of considering the QGP response, and the dashed line represents the result in the vacuum. The relation between the response time and the collision energy is shown in the subgraph. }
\end{figure}

\begin{figure}
\includegraphics[width=8.5cm]{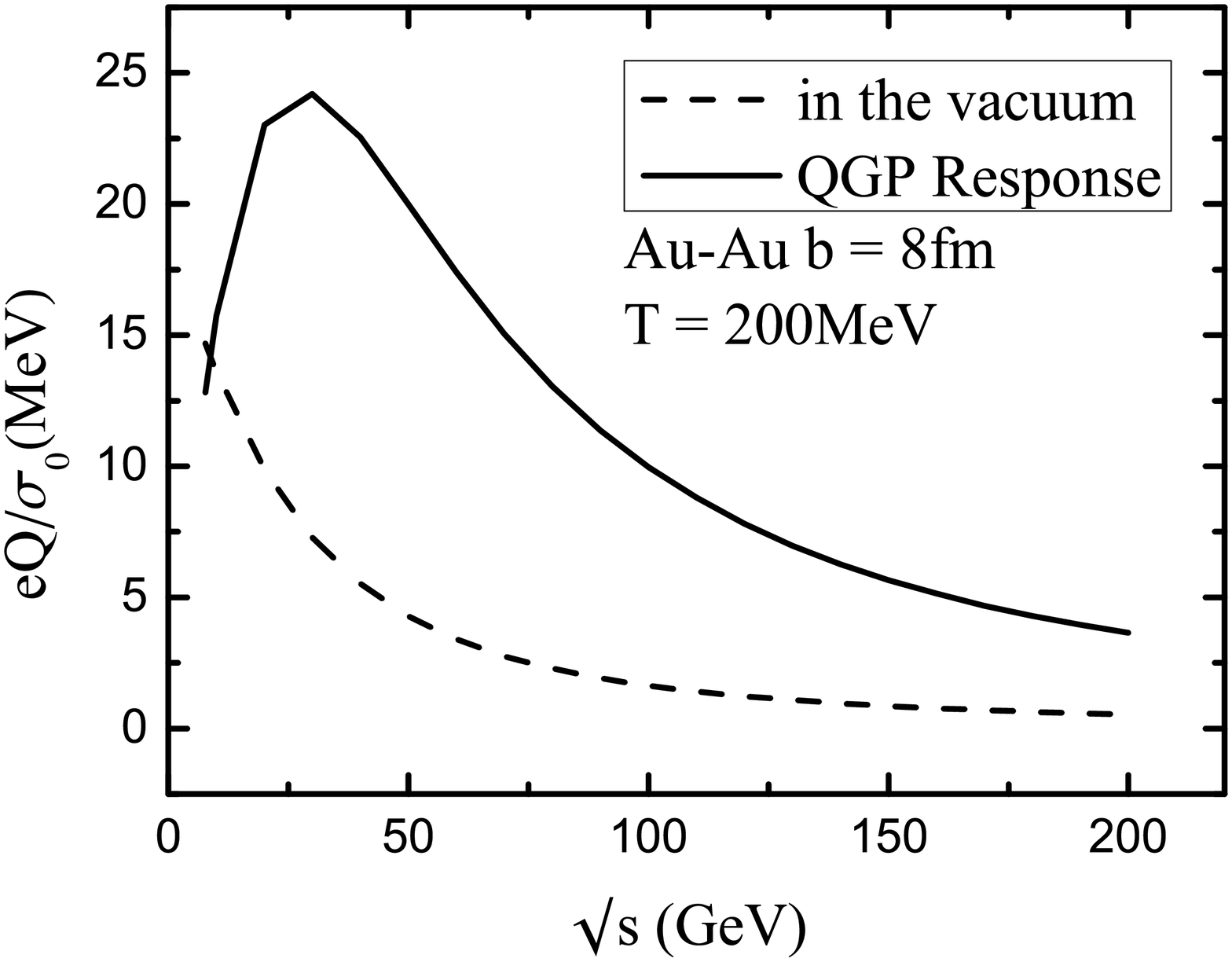}
\caption{\label{fig10}The dependence of the time-integrated current signal  ($Q = \int j(t)dt$) on the central-of-mass energy for  Au-Au collisions with $T = 200 \mathrm{MeV}$  and $b=8 \mathrm{fm}$, the solid line represents the  result of considering the QGP response, and the dashed line represents the result in the vacuum.}
\end{figure}

In order to study the dependencies of electromagnetic current on the impact parameter ($b$), we calculate
the energy dependence of formation time $t_{0}$  and magnetic field $eB^{0}_{y}(0)$ at $t=t_{0}$ at central point
$\vb{r}=\vb{0}$ for Au-Au collisions as shown in table 2.

\begin{table}
\caption{\label{tab:CuRHICdata}Impact parameter dependence of formation time $t_{0}$  and  magnetic field $eB^{0}_{y}(0)$ at $t=t_{0}$ for $\sqrt{s}=200 \mathrm{GeV}$ Au-Au collisions.}
\begin{ruledtabular}
\begin{tabular}{lll}
 $b (\mathrm{fm}) $  & $t_0 (\mathrm{fm})$ & $eB_y^0 (\mathrm{MeV}^2) $\\
 \hline
1 & 0.131 & $8.74 \times 10^{2}$  \\
2 & 0.132 & $1.71 \times 10^{3}$  \\
3 &  0.133 & $2.47 \times 10^{3}$  \\
4 & 0.135 & $3.11 \times 10^{3}$ \\
5 &  0.138  & $3.59 \times 10^{3}$ \\
6 & 0.142 &  $3.91 \times 10^{3}$ \\
7 & 0.147 & $4.00 \times 10^{3}$ \\
8 &  0.155 & $3.89 \times 10^{3}$  \\
9 & 0.167 & $3.49 \times 10^{3}$ \\
10 & 0.184 & $2.89 \times 10^{3}$ \\
11 & 0.214 &$2.05 \times 10^{3}$ \\
\end{tabular}
\end{ruledtabular}
\end{table}

Figure 11 shows the dependence of the maximum electromagnetic current on the impact parameter. When considering the QGP response, one finds that the response time is larger than that in the vacuum. The maximum electromagnetic current shows no monotonic dependence on the impact parameter. It is found that the maximum electromagnetic current $j_{max}$ increases first and reaches its maximum at $b\approx 8 \mathrm{fm}$, then decreases with the increase of the impact parameter when considering the QGP response. For the centrality dependence shown in fig.11, what is the reason behind such dependence? An important point made in Ref.\cite{Bloczynski:2012en} and the azimuthal fluctuations bring important change for the centrality dependence. our discussion is consistent with that of Ref.\cite{Bloczynski:2012en}.

\begin{figure}
\includegraphics[width=8.5cm]{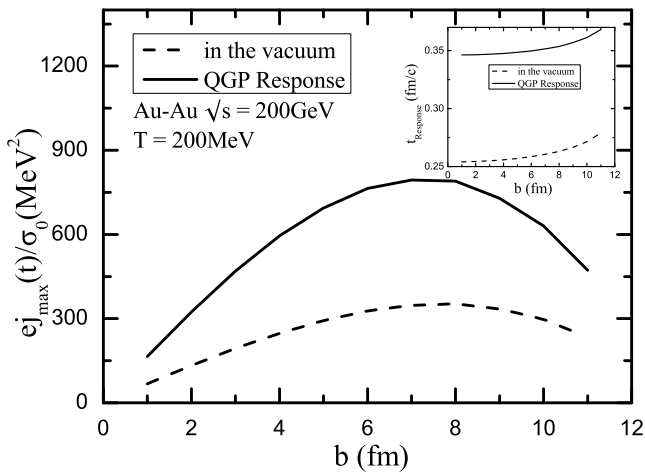}
\caption{\label{fig11}The dependence of the maximum electromagnetic current on the impact parameter $b$ for RHIC $\sqrt{s}=200 \mathrm{GeV}$ Au-Au collisions with $T = 200 \mathrm{MeV}$ , the solid line represents the  result of considering the QGP response, and the dashed line represents the result in the vacuum. The relation between the response time and impact parameter $b$ is shown in the subgraph. }
\end{figure}

Figure 12 shows the dependence of the maximum electromagnetic current on the temperature. Figure 12 indicates that the maximum electromagnetic current increases monotonically with the increase of the temperature, but the response time decreases monotonically with the increase of the temperature. The relationships among the maximum current and response time with the temperature in the vacuum are also shown in fig.12.

\begin{figure}
\includegraphics[width=8.5cm]{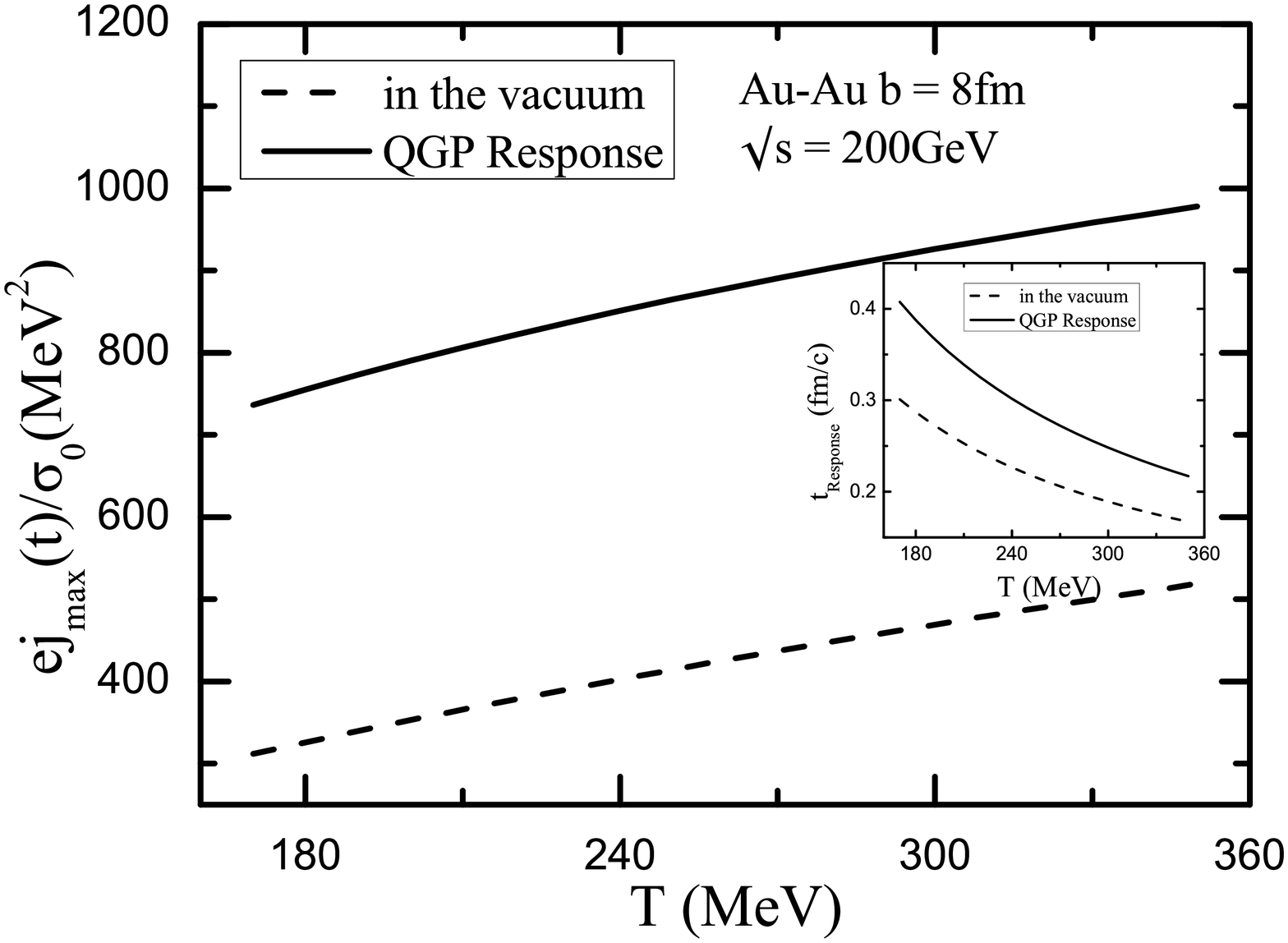}
\caption{\label{fig12}The dependence of the maximum electromagnetic current on the temperature for RHIC $\sqrt{s}=200 \mathrm{GeV}$ Au-Au collisions with $b = 8 \mathrm{fm}$ , the solid line represents the result of considering the QGP response, and the dashed line represents the result in the vacuum. The relation between the response time and the temperature is shown in the subgraph. }
\end{figure}

\section{Summary}\label{summary}
It is argued that nonzero chirality by gluon configurations with nonzero topological charge can be generated in a quark gluon plasma. In the presence of a strong magnetic field, an electromagnetic current could be produced along the field by nonzero chirality. This is the chiral magnetic effect which can potentially lead to observable effects in relativistic heavy ion collisions. Since the QGP medium is an electrically conducting medium, it is desirable to study the chiral magnetic effect in a time-dependent magnetic field by considering the response of the QGP medium.In this paper, we precisely aim to address the following two pressing issues: (1) the time evolution of the strong magnetic field which is the necessary driving force for CME, (2) the dynamical generation of CME current in response to time-dependent magnetic field. We chose the simplified model approach  to  gain valuable insights on these problems.

By considering the QGP response, we systematically study the dependencies of the electromagnetic current on temperature, collision energy and impact parameter in the RHIC and LHC energy regions.
It is found that the electromagnetic current in the LHC energy region is so small that it is difficult to produce CME. The main reason for such a small electromagnetic current in the LHC energy region is due to the sharp decrease of the magnetic field with time in the LHC energy region. Such observation is important for understanding why the recent CMS measurements \cite{Sirunyan:2017quh} at LHC $\sqrt{s}= 5020 \mathrm{GeV}$ see no CME signal.
On the other hand, the initial magnetic field ($B(t = 0)$)in the RHIC region is smaller than LHC, but the magnetic field time decreases slower with the time evolution, leading to an electromagnetic current which is much larger than that of LHC.

As we further our study of the CME research in the RHIC energy region, we find that the strongest electromagnetic current signal is not at the top RHIC energy $\sqrt{s}=200 \mathrm{GeV}$  but at collision energy $\sqrt{s} \sim 30 \mathrm{GeV}$  for Au-Au collisions at $b = 8 \mathrm{fm}$. We find that with the increase of collision energy(as $\sqrt{s} > 30 \mathrm{GeV}$ ), the corresponding electromagnetic currents become smaller and smaller, which is consistent with  some experimental studies.

\section*{Acknowledgments}
This work was supported by National Natural Science Foundation of China (Grant Nos. 11747115, 11475068), the CCNU-QLPL Innovation Fund (Grant No. QLPL2016P01), the Excellent Youth Foundation of Hubei Scientific Committee (Grant No. 2006ABB036).

\bibliography{ref}

\begin{thebibliography}{49}%
\makeatletter
\providecommand \@ifxundefined [1]{%
 \@ifx{#1\undefined}
}%
\providecommand \@ifnum [1]{%
 \ifnum #1\expandafter \@firstoftwo
 \else \expandafter \@secondoftwo
 \fi
}%
\providecommand \@ifx [1]{%
 \ifx #1\expandafter \@firstoftwo
 \else \expandafter \@secondoftwo
 \fi
}%
\providecommand \natexlab [1]{#1}%
\providecommand \enquote  [1]{``#1''}%
\providecommand \bibnamefont  [1]{#1}%
\providecommand \bibfnamefont [1]{#1}%
\providecommand \citenamefont [1]{#1}%
\providecommand \href@noop [0]{\@secondoftwo}%
\providecommand \href [0]{\begingroup \@sanitize@url \@href}%
\providecommand \@href[1]{\@@startlink{#1}\@@href}%
\providecommand \@@href[1]{\endgroup#1\@@endlink}%
\providecommand \@sanitize@url [0]{\catcode `\\12\catcode `\$12\catcode
  `\&12\catcode `\#12\catcode `\^12\catcode `\_12\catcode `\%12\relax}%
\providecommand \@@startlink[1]{}%
\providecommand \@@endlink[0]{}%
\providecommand \url  [0]{\begingroup\@sanitize@url \@url }%
\providecommand \@url [1]{\endgroup\@href {#1}{\urlprefix }}%
\providecommand \urlprefix  [0]{URL }%
\providecommand \Eprint [0]{\href }%
\providecommand \doibase [0]{http://dx.doi.org/}%
\providecommand \selectlanguage [0]{\@gobble}%
\providecommand \bibinfo  [0]{\@secondoftwo}%
\providecommand \bibfield  [0]{\@secondoftwo}%
\providecommand \translation [1]{[#1]}%
\providecommand \BibitemOpen [0]{}%
\providecommand \bibitemStop [0]{}%
\providecommand \bibitemNoStop [0]{.\EOS\space}%
\providecommand \EOS [0]{\spacefactor3000\relax}%
\providecommand \BibitemShut  [1]{\csname bibitem#1\endcsname}%
\let\auto@bib@innerbib\@empty
\bibitem [{\citenamefont {Bell}\ and\ \citenamefont
  {Jackiw}(1969)}]{Bell:1969ts}%
  \BibitemOpen
  \bibfield  {author} {\bibinfo {author} {\bibfnamefont {J.~S.}\ \bibnamefont
  {Bell}}\ and\ \bibinfo {author} {\bibfnamefont {R.}~\bibnamefont {Jackiw}},\
  }\href {\doibase 10.1007/BF02823296} {\bibfield  {journal} {\bibinfo
  {journal} {Nuovo Cim.}\ }\textbf {\bibinfo {volume} {A60}},\ \bibinfo {pages}
  {47} (\bibinfo {year} {1969})}\BibitemShut {NoStop}%
\bibitem [{\citenamefont {Adler}(1969)}]{Adler:1969gk}%
  \BibitemOpen
  \bibfield  {author} {\bibinfo {author} {\bibfnamefont {S.~L.}\ \bibnamefont
  {Adler}},\ }\href {\doibase 10.1103/PhysRev.177.2426} {\bibfield  {journal}
  {\bibinfo  {journal} {Phys. Rev.}\ }\textbf {\bibinfo {volume} {177}},\
  \bibinfo {pages} {2426} (\bibinfo {year} {1969})}\BibitemShut {NoStop}%
\bibitem [{\citenamefont {Kharzeev}\ and\ \citenamefont
  {Warringa}(2009)}]{Kharzeev:2009pj}%
  \BibitemOpen
  \bibfield  {author} {\bibinfo {author} {\bibfnamefont {D.~E.}\ \bibnamefont
  {Kharzeev}}\ and\ \bibinfo {author} {\bibfnamefont {H.~J.}\ \bibnamefont
  {Warringa}},\ }\href {\doibase 10.1103/PhysRevD.80.034028} {\bibfield
  {journal} {\bibinfo  {journal} {Phys. Rev.}\ }\textbf {\bibinfo {volume}
  {D80}},\ \bibinfo {pages} {034028} (\bibinfo {year} {2009})}\BibitemShut
  {NoStop}%
\bibitem [{\citenamefont {Fukushima}\ \emph {et~al.}(2008)\citenamefont
  {Fukushima}, \citenamefont {Kharzeev},\ and\ \citenamefont
  {Warringa}}]{Fukushima:2008xe}%
  \BibitemOpen
  \bibfield  {author} {\bibinfo {author} {\bibfnamefont {K.}~\bibnamefont
  {Fukushima}}, \bibinfo {author} {\bibfnamefont {D.~E.}\ \bibnamefont
  {Kharzeev}}, \ and\ \bibinfo {author} {\bibfnamefont {H.~J.}\ \bibnamefont
  {Warringa}},\ }\href {\doibase 10.1103/PhysRevD.78.074033} {\bibfield
  {journal} {\bibinfo  {journal} {Phys. Rev. D}\ }\textbf {\bibinfo {volume}
  {78}},\ \bibinfo {pages} {074033} (\bibinfo {year} {2008})}\BibitemShut
  {NoStop}%
\bibitem [{\citenamefont {Kharzeev}\ \emph {et~al.}(2008)\citenamefont
  {Kharzeev}, \citenamefont {McLerran},\ and\ \citenamefont
  {Warringa}}]{Kharzeev:2007jp}%
  \BibitemOpen
  \bibfield  {author} {\bibinfo {author} {\bibfnamefont {D.~E.}\ \bibnamefont
  {Kharzeev}}, \bibinfo {author} {\bibfnamefont {L.~D.}\ \bibnamefont
  {McLerran}}, \ and\ \bibinfo {author} {\bibfnamefont {H.~J.}\ \bibnamefont
  {Warringa}},\ }\href {\doibase 10.1016/j.nuclphysa.2008.02.298} {\bibfield
  {journal} {\bibinfo  {journal} {Nucl. Phys. A}\ }\textbf {\bibinfo {volume}
  {803}},\ \bibinfo {pages} {227} (\bibinfo {year} {2008})}\BibitemShut
  {NoStop}%
\bibitem [{\citenamefont {Kharzeev}\ and\ \citenamefont
  {Zhitnitsky}(2007)}]{Kharzeev:2007tn}%
  \BibitemOpen
  \bibfield  {author} {\bibinfo {author} {\bibfnamefont {D.}~\bibnamefont
  {Kharzeev}}\ and\ \bibinfo {author} {\bibfnamefont {A.}~\bibnamefont
  {Zhitnitsky}},\ }\href {\doibase 10.1016/j.nuclphysa.2007.10.001} {\bibfield
  {journal} {\bibinfo  {journal} {Nucl. Phys.}\ }\textbf {\bibinfo {volume}
  {A797}},\ \bibinfo {pages} {67} (\bibinfo {year} {2007})}\BibitemShut
  {NoStop}%
\bibitem [{\citenamefont {Kharzeev}(2006)}]{Kharzeev:2004ey}%
  \BibitemOpen
  \bibfield  {author} {\bibinfo {author} {\bibfnamefont {D.}~\bibnamefont
  {Kharzeev}},\ }\href {\doibase 10.1016/j.physletb.2005.11.075} {\bibfield
  {journal} {\bibinfo  {journal} {Phys. Lett.}\ }\textbf {\bibinfo {volume}
  {B633}},\ \bibinfo {pages} {260} (\bibinfo {year} {2006})}\BibitemShut
  {NoStop}%
\bibitem [{\citenamefont {Burnier}\ \emph {et~al.}(2011)\citenamefont
  {Burnier}, \citenamefont {Kharzeev}, \citenamefont {Liao},\ and\
  \citenamefont {Yee}}]{Burnier:2011bf}%
  \BibitemOpen
  \bibfield  {author} {\bibinfo {author} {\bibfnamefont {Y.}~\bibnamefont
  {Burnier}}, \bibinfo {author} {\bibfnamefont {D.~E.}\ \bibnamefont
  {Kharzeev}}, \bibinfo {author} {\bibfnamefont {J.}~\bibnamefont {Liao}}, \
  and\ \bibinfo {author} {\bibfnamefont {H.-U.}\ \bibnamefont {Yee}},\ }\href
  {\doibase 10.1103/PhysRevLett.107.052303} {\bibfield  {journal} {\bibinfo
  {journal} {Phys. Rev. Lett.}\ }\textbf {\bibinfo {volume} {107}},\ \bibinfo
  {pages} {052303} (\bibinfo {year} {2011})}\BibitemShut {NoStop}%
\bibitem [{\citenamefont {Mo}\ \emph {et~al.}(2013)\citenamefont {Mo},
  \citenamefont {Feng},\ and\ \citenamefont {Shi}}]{Mo:2013qya}%
  \BibitemOpen
  \bibfield  {author} {\bibinfo {author} {\bibfnamefont {Y.-J.}\ \bibnamefont
  {Mo}}, \bibinfo {author} {\bibfnamefont {S.-Q.}\ \bibnamefont {Feng}}, \ and\
  \bibinfo {author} {\bibfnamefont {Y.-F.}\ \bibnamefont {Shi}},\ }\href
  {\doibase 10.1103/PhysRevC.88.024901} {\bibfield  {journal} {\bibinfo
  {journal} {Phys. Rev. C}\ }\textbf {\bibinfo {volume} {88}},\ \bibinfo
  {pages} {024901} (\bibinfo {year} {2013})}\BibitemShut {NoStop}%
\bibitem [{\citenamefont {Skokov}\ \emph {et~al.}(2009)\citenamefont {Skokov},
  \citenamefont {Illarionov},\ and\ \citenamefont {Toneev}}]{Skokov:2009qp}%
  \BibitemOpen
  \bibfield  {author} {\bibinfo {author} {\bibfnamefont {V.}~\bibnamefont
  {Skokov}}, \bibinfo {author} {\bibfnamefont {A.~{\relax Yu}.}\ \bibnamefont
  {Illarionov}}, \ and\ \bibinfo {author} {\bibfnamefont {V.}~\bibnamefont
  {Toneev}},\ }\href {\doibase 10.1142/S0217751X09047570} {\bibfield  {journal}
  {\bibinfo  {journal} {Int. J. Mod. Phys. A}\ }\textbf {\bibinfo {volume}
  {24}},\ \bibinfo {pages} {5925} (\bibinfo {year} {2009})}\BibitemShut
  {NoStop}%
\bibitem [{\citenamefont {Voronyuk}\ \emph {et~al.}(2011)\citenamefont
  {Voronyuk}, \citenamefont {Toneev}, \citenamefont {Cassing}, \citenamefont
  {Bratkovskaya}, \citenamefont {Konchakovski},\ and\ \citenamefont
  {Voloshin}}]{Voronyuk:2011jd}%
  \BibitemOpen
  \bibfield  {author} {\bibinfo {author} {\bibfnamefont {V.}~\bibnamefont
  {Voronyuk}}, \bibinfo {author} {\bibfnamefont {V.~D.}\ \bibnamefont
  {Toneev}}, \bibinfo {author} {\bibfnamefont {W.}~\bibnamefont {Cassing}},
  \bibinfo {author} {\bibfnamefont {E.~L.}\ \bibnamefont {Bratkovskaya}},
  \bibinfo {author} {\bibfnamefont {V.~P.}\ \bibnamefont {Konchakovski}}, \
  and\ \bibinfo {author} {\bibfnamefont {S.~A.}\ \bibnamefont {Voloshin}},\
  }\href {\doibase 10.1103/PhysRevC.83.054911} {\bibfield  {journal} {\bibinfo
  {journal} {Phys. Rev. C}\ }\textbf {\bibinfo {volume} {83}},\ \bibinfo
  {pages} {054911} (\bibinfo {year} {2011})}\BibitemShut {NoStop}%
\bibitem [{\citenamefont {Bzdak}\ and\ \citenamefont
  {Skokov}(2012)}]{Bzdak:2011yy}%
  \BibitemOpen
  \bibfield  {author} {\bibinfo {author} {\bibfnamefont {A.}~\bibnamefont
  {Bzdak}}\ and\ \bibinfo {author} {\bibfnamefont {V.}~\bibnamefont {Skokov}},\
  }\href {\doibase 10.1016/j.physletb.2012.02.065} {\bibfield  {journal}
  {\bibinfo  {journal} {Phys. Lett. B}\ }\textbf {\bibinfo {volume} {710}},\
  \bibinfo {pages} {171} (\bibinfo {year} {2012})}\BibitemShut {NoStop}%
\bibitem [{\citenamefont {Tuchin}(2013{\natexlab{a}})}]{Tuchin:2013apa}%
  \BibitemOpen
  \bibfield  {author} {\bibinfo {author} {\bibfnamefont {K.}~\bibnamefont
  {Tuchin}},\ }\href {\doibase 10.1103/PhysRevC.88.024911} {\bibfield
  {journal} {\bibinfo  {journal} {Phys. Rev. C}\ }\textbf {\bibinfo {volume}
  {88}},\ \bibinfo {pages} {024911} (\bibinfo {year}
  {2013}{\natexlab{a}})}\BibitemShut {NoStop}%
\bibitem [{\citenamefont {Deng}\ and\ \citenamefont
  {Huang}(2012)}]{Deng:2012pc}%
  \BibitemOpen
  \bibfield  {author} {\bibinfo {author} {\bibfnamefont {W.-T.}\ \bibnamefont
  {Deng}}\ and\ \bibinfo {author} {\bibfnamefont {X.-G.}\ \bibnamefont
  {Huang}},\ }\href {\doibase 10.1103/PhysRevC.85.044907} {\bibfield  {journal}
  {\bibinfo  {journal} {Phys. Rev. C}\ }\textbf {\bibinfo {volume} {85}},\
  \bibinfo {pages} {044907} (\bibinfo {year} {2012})}\BibitemShut {NoStop}%
\bibitem [{\citenamefont {Wang}\ and\ \citenamefont
  {Wen}(2017)}]{Wang:2016mkm}%
  \BibitemOpen
  \bibfield  {author} {\bibinfo {author} {\bibfnamefont {G.}~\bibnamefont
  {Wang}}\ and\ \bibinfo {author} {\bibfnamefont {L.}~\bibnamefont {Wen}},\
  }\href {\doibase 10.1155/2017/9240170} {\bibfield  {journal} {\bibinfo
  {journal} {Adv. High Energy Phys.}\ }\textbf {\bibinfo {volume} {2017}},\
  \bibinfo {pages} {9240170} (\bibinfo {year} {2017})}\BibitemShut {NoStop}%
\bibitem [{\citenamefont {Adamczyk}\ \emph
  {et~al.}(2014{\natexlab{a}})\citenamefont {Adamczyk} \emph
  {et~al.}}]{Adamczyk:2013kcb}%
  \BibitemOpen
  \bibfield  {author} {\bibinfo {author} {\bibfnamefont {L.}~\bibnamefont
  {Adamczyk}} \emph {et~al.} (\bibinfo {collaboration} {STAR Collaboration}),\
  }\href {\doibase 10.1103/PhysRevC.89.044908} {\bibfield  {journal} {\bibinfo
  {journal} {Phys. Rev. C}\ }\textbf {\bibinfo {volume} {89}},\ \bibinfo
  {pages} {044908} (\bibinfo {year} {2014}{\natexlab{a}})}\BibitemShut
  {NoStop}%
\bibitem [{\citenamefont {Adamczyk}\ \emph
  {et~al.}(2014{\natexlab{b}})\citenamefont {Adamczyk} \emph
  {et~al.}}]{Adamczyk:2014mzf}%
  \BibitemOpen
  \bibfield  {author} {\bibinfo {author} {\bibfnamefont {L.}~\bibnamefont
  {Adamczyk}} \emph {et~al.} (\bibinfo {collaboration} {STAR Collaboration}),\
  }\href {\doibase 10.1103/PhysRevLett.113.052302} {\bibfield  {journal}
  {\bibinfo  {journal} {Phys. Rev. Lett.}\ }\textbf {\bibinfo {volume} {113}},\
  \bibinfo {pages} {052302} (\bibinfo {year} {2014}{\natexlab{b}})}\BibitemShut
  {NoStop}%
\bibitem [{\citenamefont {Wang}(2013)}]{Wang:2012qs}%
  \BibitemOpen
  \bibfield  {author} {\bibinfo {author} {\bibfnamefont {G.}~\bibnamefont
  {Wang}} (\bibinfo {collaboration} {STAR Collaboration}),\ }\href {\doibase
  10.1016/j.nuclphysa.2013.01.069} {\bibfield  {journal} {\bibinfo  {journal}
  {Nucl. Phys. A}\ }\textbf {\bibinfo {volume} {904-905}},\ \bibinfo {pages}
  {248c} (\bibinfo {year} {2013})}\BibitemShut {NoStop}%
\bibitem [{\citenamefont {Adamczyk}\ \emph {et~al.}(2013)\citenamefont
  {Adamczyk} \emph {et~al.}}]{PhysRevC.88.064911}%
  \BibitemOpen
  \bibfield  {author} {\bibinfo {author} {\bibfnamefont {L.}~\bibnamefont
  {Adamczyk}} \emph {et~al.} (\bibinfo {collaboration} {STAR Collaboration}),\
  }\href {\doibase 10.1103/PhysRevC.88.064911} {\bibfield  {journal} {\bibinfo
  {journal} {Phys. Rev. C}\ }\textbf {\bibinfo {volume} {88}},\ \bibinfo
  {pages} {064911} (\bibinfo {year} {2013})}\BibitemShut {NoStop}%
\bibitem [{\citenamefont {Voloshin}\ \emph {et~al.}(2011)\citenamefont
  {Voloshin} \emph {et~al.}}]{Voloshin:2008jx}%
  \BibitemOpen
  \bibfield  {author} {\bibinfo {author} {\bibfnamefont {S.~A.}\ \bibnamefont
  {Voloshin}} \emph {et~al.} (\bibinfo {collaboration} {STAR Collaboration}),\
  }\href {\doibase 10.1007/s12648-011-0137-0} {\bibfield  {journal} {\bibinfo
  {journal} {Indian J. Phys.}\ }\textbf {\bibinfo {volume} {85}},\ \bibinfo
  {pages} {1103} (\bibinfo {year} {2011})}\BibitemShut {NoStop}%
\bibitem [{\citenamefont {Abelev}\ \emph {et~al.}(2010)\citenamefont {Abelev}
  \emph {et~al.}}]{Abelev:2009ad}%
  \BibitemOpen
  \bibfield  {author} {\bibinfo {author} {\bibfnamefont {B.~I.}\ \bibnamefont
  {Abelev}} \emph {et~al.} (\bibinfo {collaboration} {STAR Collaboration}),\
  }\href {\doibase 10.1103/PhysRevC.81.054908} {\bibfield  {journal} {\bibinfo
  {journal} {Phys. Rev. C}\ }\textbf {\bibinfo {volume} {81}},\ \bibinfo
  {pages} {054908} (\bibinfo {year} {2010})}\BibitemShut {NoStop}%
\bibitem [{\citenamefont {Abelev}\ \emph {et~al.}(2009)\citenamefont {Abelev}
  \emph {et~al.}}]{Abelev2009ac}%
  \BibitemOpen
  \bibfield  {author} {\bibinfo {author} {\bibfnamefont {B.~I.}\ \bibnamefont
  {Abelev}} \emph {et~al.} (\bibinfo {collaboration} {STAR Collaboration}),\
  }\href {\doibase 10.1103/PhysRevLett.103.251601} {\bibfield  {journal}
  {\bibinfo  {journal} {Phys. Rev. Lett.}\ }\textbf {\bibinfo {volume} {103}},\
  \bibinfo {pages} {251601} (\bibinfo {year} {2009})}\BibitemShut {NoStop}%
\bibitem [{\citenamefont {Abelev}\ \emph {et~al.}(2013)\citenamefont {Abelev}
  \emph {et~al.}}]{PhysRevLett.110.012301}%
  \BibitemOpen
  \bibfield  {author} {\bibinfo {author} {\bibfnamefont {B.}~\bibnamefont
  {Abelev}} \emph {et~al.} (\bibinfo {collaboration} {ALICE Collaboration}),\
  }\href {\doibase 10.1103/PhysRevLett.110.012301} {\bibfield  {journal}
  {\bibinfo  {journal} {Phys. Rev. Lett.}\ }\textbf {\bibinfo {volume} {110}},\
  \bibinfo {pages} {012301} (\bibinfo {year} {2013})}\BibitemShut {NoStop}%
\bibitem [{\citenamefont {Christakoglou}(2011)}]{Christakoglou:2011uqg}%
  \BibitemOpen
  \bibfield  {author} {\bibinfo {author} {\bibfnamefont {P.}~\bibnamefont
  {Christakoglou}} (\bibinfo {collaboration} {ALICE Collaboration}),\ }\href
  {\doibase 10.1088/0954-3899/38/12/124165} {\bibfield  {journal} {\bibinfo
  {journal} {J. Phys. G}\ }\textbf {\bibinfo {volume} {38}},\ \bibinfo {pages}
  {124165} (\bibinfo {year} {2011})}\BibitemShut {NoStop}%
\bibitem [{\citenamefont {Kharzeev}(2014)}]{Kharzeev:2013ffa}%
  \BibitemOpen
  \bibfield  {author} {\bibinfo {author} {\bibfnamefont {D.~E.}\ \bibnamefont
  {Kharzeev}},\ }\href {\doibase 10.1016/j.ppnp.2014.01.002} {\bibfield
  {journal} {\bibinfo  {journal} {Prog. Part. Nucl. Phys.}\ }\textbf {\bibinfo
  {volume} {75}},\ \bibinfo {pages} {133} (\bibinfo {year} {2014})}\BibitemShut
  {NoStop}%
\bibitem [{\citenamefont {Kharzeev}\ \emph {et~al.}(2016)\citenamefont
  {Kharzeev}, \citenamefont {Liao}, \citenamefont {Voloshin},\ and\
  \citenamefont {Wang}}]{Kharzeev:2015znc}%
  \BibitemOpen
  \bibfield  {author} {\bibinfo {author} {\bibfnamefont {D.~E.}\ \bibnamefont
  {Kharzeev}}, \bibinfo {author} {\bibfnamefont {J.}~\bibnamefont {Liao}},
  \bibinfo {author} {\bibfnamefont {S.~A.}\ \bibnamefont {Voloshin}}, \ and\
  \bibinfo {author} {\bibfnamefont {G.}~\bibnamefont {Wang}},\ }\href {\doibase
  10.1016/j.ppnp.2016.01.001} {\bibfield  {journal} {\bibinfo  {journal} {Prog.
  Part. Nucl. Phys.}\ }\textbf {\bibinfo {volume} {88}},\ \bibinfo {pages} {1}
  (\bibinfo {year} {2016})}\BibitemShut {NoStop}%
\bibitem [{\citenamefont {Tuchin}(2010)}]{Tuchin:2010vs}%
  \BibitemOpen
  \bibfield  {author} {\bibinfo {author} {\bibfnamefont {K.}~\bibnamefont
  {Tuchin}},\ }\href {\doibase 10.1103/PhysRevC.82.034904} {\bibfield
  {journal} {\bibinfo  {journal} {Phys. Rev. C}\ }\textbf {\bibinfo {volume}
  {82}},\ \bibinfo {pages} {034904} (\bibinfo {year} {2010})}\BibitemShut
  {NoStop}%
\bibitem [{\citenamefont {Tuchin}(2011)}]{Tuchin:2010vs2}%
  \BibitemOpen
  \bibfield  {author} {\bibinfo {author} {\bibfnamefont {K.}~\bibnamefont
  {Tuchin}},\ }\href {\doibase 10.1103/PhysRevC.83.039903} {\bibfield
  {journal} {\bibinfo  {journal} {Phys. Rev. C}\ }\textbf {\bibinfo {volume}
  {83}},\ \bibinfo {pages} {039903} (\bibinfo {year} {2011})},\ \bibinfo {note}
  {[Erratum]}\BibitemShut {NoStop}%
\bibitem [{\citenamefont {McLerran}\ and\ \citenamefont
  {Skokov}(2014)}]{McLerran:2013hla}%
  \BibitemOpen
  \bibfield  {author} {\bibinfo {author} {\bibfnamefont {L.}~\bibnamefont
  {McLerran}}\ and\ \bibinfo {author} {\bibfnamefont {V.}~\bibnamefont
  {Skokov}},\ }\href {\doibase 10.1016/j.nuclphysa.2014.05.008} {\bibfield
  {journal} {\bibinfo  {journal} {Nucl. Phys. A}\ }\textbf {\bibinfo {volume}
  {929}},\ \bibinfo {pages} {184} (\bibinfo {year} {2014})}\BibitemShut
  {NoStop}%
\bibitem [{\citenamefont {Tuchin}(2013{\natexlab{b}})}]{Tuchin:2013ie}%
  \BibitemOpen
  \bibfield  {author} {\bibinfo {author} {\bibfnamefont {K.}~\bibnamefont
  {Tuchin}},\ }\href {\doibase 10.1155/2013/490495} {\bibfield  {journal}
  {\bibinfo  {journal} {Adv. High Energy Phys.}\ }\textbf {\bibinfo {volume}
  {2013}},\ \bibinfo {pages} {490495} (\bibinfo {year}
  {2013}{\natexlab{b}})}\BibitemShut {NoStop}%
\bibitem [{\citenamefont {Zakharov}(2014)}]{Zakharov:2014dia}%
  \BibitemOpen
  \bibfield  {author} {\bibinfo {author} {\bibfnamefont {B.~G.}\ \bibnamefont
  {Zakharov}},\ }\href {\doibase 10.1016/j.physletb.2014.08.068} {\bibfield
  {journal} {\bibinfo  {journal} {Phys. Lett. B}\ }\textbf {\bibinfo {volume}
  {737}},\ \bibinfo {pages} {262} (\bibinfo {year} {2014})}\BibitemShut
  {NoStop}%
\bibitem [{\citenamefont {Tuchin}(2015)}]{Tuchin:2014iua}%
  \BibitemOpen
  \bibfield  {author} {\bibinfo {author} {\bibfnamefont {K.}~\bibnamefont
  {Tuchin}},\ }\href {\doibase 10.1103/PhysRevC.91.064902} {\bibfield
  {journal} {\bibinfo  {journal} {Phys. Rev. C}\ }\textbf {\bibinfo {volume}
  {91}},\ \bibinfo {pages} {064902} (\bibinfo {year} {2015})}\BibitemShut
  {NoStop}%
\bibitem [{\citenamefont {Tuchin}(2016)}]{Tuchin:2015oka}%
  \BibitemOpen
  \bibfield  {author} {\bibinfo {author} {\bibfnamefont {K.}~\bibnamefont
  {Tuchin}},\ }\href {\doibase 10.1103/PhysRevC.93.014905} {\bibfield
  {journal} {\bibinfo  {journal} {Phys. Rev. C}\ }\textbf {\bibinfo {volume}
  {93}},\ \bibinfo {pages} {014905} (\bibinfo {year} {2016})}\BibitemShut
  {NoStop}%
\bibitem [{\citenamefont {Francis}\ and\ \citenamefont
  {Kaczmarek}(2012)}]{Francis:2011bt}%
  \BibitemOpen
  \bibfield  {author} {\bibinfo {author} {\bibfnamefont {A.}~\bibnamefont
  {Francis}}\ and\ \bibinfo {author} {\bibfnamefont {O.}~\bibnamefont
  {Kaczmarek}},\ }\href {\doibase 10.1016/j.ppnp.2011.12.020} {\bibfield
  {journal} {\bibinfo  {journal} {Prog. Part. Nucl. Phys.}\ }\textbf {\bibinfo
  {volume} {67}},\ \bibinfo {pages} {212} (\bibinfo {year} {2012})}\BibitemShut
  {NoStop}%
\bibitem [{\citenamefont {Ding}\ \emph {et~al.}(2011)\citenamefont {Ding},
  \citenamefont {Francis}, \citenamefont {Kaczmarek}, \citenamefont {Karsch},
  \citenamefont {Laermann},\ and\ \citenamefont {Soeldner}}]{Ding:2010ga}%
  \BibitemOpen
  \bibfield  {author} {\bibinfo {author} {\bibfnamefont {H.~T.}\ \bibnamefont
  {Ding}}, \bibinfo {author} {\bibfnamefont {A.}~\bibnamefont {Francis}},
  \bibinfo {author} {\bibfnamefont {O.}~\bibnamefont {Kaczmarek}}, \bibinfo
  {author} {\bibfnamefont {F.}~\bibnamefont {Karsch}}, \bibinfo {author}
  {\bibfnamefont {E.}~\bibnamefont {Laermann}}, \ and\ \bibinfo {author}
  {\bibfnamefont {W.}~\bibnamefont {Soeldner}},\ }\href {\doibase
  10.1103/PhysRevD.83.034504} {\bibfield  {journal} {\bibinfo  {journal} {Phys.
  Rev. D}\ }\textbf {\bibinfo {volume} {83}},\ \bibinfo {pages} {034504}
  (\bibinfo {year} {2011})}\BibitemShut {NoStop}%
\bibitem [{\citenamefont {Arnold}\ \emph {et~al.}(2003)\citenamefont {Arnold},
  \citenamefont {Moore},\ and\ \citenamefont {Yaffe}}]{Arnold:2003zc}%
  \BibitemOpen
  \bibfield  {author} {\bibinfo {author} {\bibfnamefont {P.~B.}\ \bibnamefont
  {Arnold}}, \bibinfo {author} {\bibfnamefont {G.~D.}\ \bibnamefont {Moore}}, \
  and\ \bibinfo {author} {\bibfnamefont {L.~G.}\ \bibnamefont {Yaffe}},\ }\href
  {\doibase 10.1088/1126-6708/2003/05/051} {\bibfield  {journal} {\bibinfo
  {journal} {JHEP}\ }\textbf {\bibinfo {volume} {05}},\ \bibinfo {pages} {051}
  (\bibinfo {year} {2003})}\BibitemShut {NoStop}%
\bibitem [{\citenamefont {Gupta}(2004)}]{Gupta:2003zh}%
  \BibitemOpen
  \bibfield  {author} {\bibinfo {author} {\bibfnamefont {S.}~\bibnamefont
  {Gupta}},\ }\href {\doibase 10.1016/j.physletb.2004.05.079} {\bibfield
  {journal} {\bibinfo  {journal} {Phys. Lett. B}\ }\textbf {\bibinfo {volume}
  {597}},\ \bibinfo {pages} {57} (\bibinfo {year} {2004})}\BibitemShut
  {NoStop}%
\bibitem [{\citenamefont {Bjorken}(1983)}]{Bjorken:1982qr}%
  \BibitemOpen
  \bibfield  {author} {\bibinfo {author} {\bibfnamefont {J.~D.}\ \bibnamefont
  {Bjorken}},\ }\href {\doibase 10.1103/PhysRevD.27.140} {\bibfield  {journal}
  {\bibinfo  {journal} {Phys. Rev.}\ }\textbf {\bibinfo {volume} {D27}},\
  \bibinfo {pages} {140} (\bibinfo {year} {1983})}\BibitemShut {NoStop}%
\bibitem [{\citenamefont {Ollitrault}(2008)}]{Ollitrault:2008zz}%
  \BibitemOpen
  \bibfield  {author} {\bibinfo {author} {\bibfnamefont {J.-Y.}\ \bibnamefont
  {Ollitrault}},\ }\href {\doibase 10.1088/0143-0807/29/2/010} {\bibfield
  {journal} {\bibinfo  {journal} {Eur. J. Phys.}\ }\textbf {\bibinfo {volume}
  {29}},\ \bibinfo {pages} {275} (\bibinfo {year} {2008})}\BibitemShut
  {NoStop}%
\bibitem [{\citenamefont {Kharzeev}\ and\ \citenamefont
  {Nardi}(2001)}]{Kharzeev:2000ph}%
  \BibitemOpen
  \bibfield  {author} {\bibinfo {author} {\bibfnamefont {D.}~\bibnamefont
  {Kharzeev}}\ and\ \bibinfo {author} {\bibfnamefont {M.}~\bibnamefont
  {Nardi}},\ }\href {\doibase 10.1016/S0370-2693(01)00457-9} {\bibfield
  {journal} {\bibinfo  {journal} {Phys. Lett. B}\ }\textbf {\bibinfo {volume}
  {507}},\ \bibinfo {pages} {121} (\bibinfo {year} {2001})}\BibitemShut
  {NoStop}%
\bibitem [{\citenamefont {Kowalski}\ \emph {et~al.}(2008)\citenamefont
  {Kowalski}, \citenamefont {Lappi},\ and\ \citenamefont
  {Venugopalan}}]{Kowalski:2007rw}%
  \BibitemOpen
  \bibfield  {author} {\bibinfo {author} {\bibfnamefont {H.}~\bibnamefont
  {Kowalski}}, \bibinfo {author} {\bibfnamefont {T.}~\bibnamefont {Lappi}}, \
  and\ \bibinfo {author} {\bibfnamefont {R.}~\bibnamefont {Venugopalan}},\
  }\href {\doibase 10.1103/PhysRevLett.100.022303} {\bibfield  {journal}
  {\bibinfo  {journal} {Phys. Rev. Lett.}\ }\textbf {\bibinfo {volume} {100}},\
  \bibinfo {pages} {022303} (\bibinfo {year} {2008})}\BibitemShut {NoStop}%
\bibitem [{\citenamefont {Lappi}(2009)}]{Lappi:2009mp}%
  \BibitemOpen
  \bibfield  {author} {\bibinfo {author} {\bibfnamefont {T.}~\bibnamefont
  {Lappi}},\ }\href {\doibase 10.1016/j.nuclphysa.2009.05.079} {\bibfield
  {journal} {\bibinfo  {journal} {Nucl. Phys.}\ }\textbf {\bibinfo {volume}
  {A827}},\ \bibinfo {pages} {365C} (\bibinfo {year} {2009})}\BibitemShut
  {NoStop}%
\bibitem [{\citenamefont {Inghirami}\ \emph {et~al.}(2016)\citenamefont
  {Inghirami}, \citenamefont {Del~Zanna}, \citenamefont {Beraudo},
  \citenamefont {Moghaddam}, \citenamefont {Becattini},\ and\ \citenamefont
  {Bleicher}}]{Inghirami:2016iru}%
  \BibitemOpen
  \bibfield  {author} {\bibinfo {author} {\bibfnamefont {G.}~\bibnamefont
  {Inghirami}}, \bibinfo {author} {\bibfnamefont {L.}~\bibnamefont
  {Del~Zanna}}, \bibinfo {author} {\bibfnamefont {A.}~\bibnamefont {Beraudo}},
  \bibinfo {author} {\bibfnamefont {M.~H.}\ \bibnamefont {Moghaddam}}, \bibinfo
  {author} {\bibfnamefont {F.}~\bibnamefont {Becattini}}, \ and\ \bibinfo
  {author} {\bibfnamefont {M.}~\bibnamefont {Bleicher}},\ }\href {\doibase
  10.1140/epjc/s10052-016-4516-8} {\bibfield  {journal} {\bibinfo  {journal}
  {Eur. Phys. J.}\ }\textbf {\bibinfo {volume} {C76}},\ \bibinfo {pages} {659}
  (\bibinfo {year} {2016})}\BibitemShut {NoStop}%
\bibitem [{\citenamefont {Jiang}\ \emph {et~al.}(2018)\citenamefont {Jiang},
  \citenamefont {Shi}, \citenamefont {Yin},\ and\ \citenamefont
  {Liao}}]{Jiang:2016wve}%
  \BibitemOpen
  \bibfield  {author} {\bibinfo {author} {\bibfnamefont {Y.}~\bibnamefont
  {Jiang}}, \bibinfo {author} {\bibfnamefont {S.}~\bibnamefont {Shi}}, \bibinfo
  {author} {\bibfnamefont {Y.}~\bibnamefont {Yin}}, \ and\ \bibinfo {author}
  {\bibfnamefont {J.}~\bibnamefont {Liao}},\ }\href {\doibase
  10.1088/1674-1137/42/1/011001} {\bibfield  {journal} {\bibinfo  {journal}
  {Chin. Phys.}\ }\textbf {\bibinfo {volume} {C42}},\ \bibinfo {pages} {011001}
  (\bibinfo {year} {2018})}\BibitemShut {NoStop}%
\bibitem [{\citenamefont {Shi}\ \emph {et~al.}(2017)\citenamefont {Shi},
  \citenamefont {Jiang}, \citenamefont {Lilleskov},\ and\ \citenamefont
  {Liao}}]{Shi:2017cpu}%
  \BibitemOpen
  \bibfield  {author} {\bibinfo {author} {\bibfnamefont {S.}~\bibnamefont
  {Shi}}, \bibinfo {author} {\bibfnamefont {Y.}~\bibnamefont {Jiang}}, \bibinfo
  {author} {\bibfnamefont {E.}~\bibnamefont {Lilleskov}}, \ and\ \bibinfo
  {author} {\bibfnamefont {J.}~\bibnamefont {Liao}},\ }\href@noop {} {\
  (\bibinfo {year} {2017})},\ \Eprint {http://arxiv.org/abs/1711.02496}
  {arXiv:1711.02496 [nucl-th]} \BibitemShut {NoStop}%
\bibitem [{\citenamefont {Sirunyan}\ \emph {et~al.}(2017)\citenamefont
  {Sirunyan} \emph {et~al.}}]{Sirunyan:2017quh}%
  \BibitemOpen
  \bibfield  {author} {\bibinfo {author} {\bibfnamefont {A.~M.}\ \bibnamefont
  {Sirunyan}} \emph {et~al.} (\bibinfo {collaboration} {CMS}),\ }\href@noop {}
  {\  (\bibinfo {year} {2017})},\ \Eprint {http://arxiv.org/abs/1708.01602}
  {arXiv:1708.01602 [nucl-ex]} \BibitemShut {NoStop}%
\bibitem [{\citenamefont {Bzdak}\ \emph {et~al.}(2013)\citenamefont {Bzdak},
  \citenamefont {Koch},\ and\ \citenamefont {Liao}}]{Bzdak:2012ia}%
  \BibitemOpen
  \bibfield  {author} {\bibinfo {author} {\bibfnamefont {A.}~\bibnamefont
  {Bzdak}}, \bibinfo {author} {\bibfnamefont {V.}~\bibnamefont {Koch}}, \ and\
  \bibinfo {author} {\bibfnamefont {J.}~\bibnamefont {Liao}},\ }\href {\doibase
  10.1007/978-3-642-37305-3_19} {\bibfield  {journal} {\bibinfo  {journal}
  {Lect. Notes Phys.}\ }\textbf {\bibinfo {volume} {871}},\ \bibinfo {pages}
  {503} (\bibinfo {year} {2013})}\BibitemShut {NoStop}%
\bibitem [{\citenamefont {Liao}(2015)}]{Liao:2014ava}%
  \BibitemOpen
  \bibfield  {author} {\bibinfo {author} {\bibfnamefont {J.}~\bibnamefont
  {Liao}},\ }\href {\doibase 10.1007/s12043-015-0984-x} {\bibfield  {journal}
  {\bibinfo  {journal} {Pramana}\ }\textbf {\bibinfo {volume} {84}},\ \bibinfo
  {pages} {901} (\bibinfo {year} {2015})}\BibitemShut {NoStop}%
\bibitem [{\citenamefont {Bloczynski}\ \emph {et~al.}(2013)\citenamefont
  {Bloczynski}, \citenamefont {Huang}, \citenamefont {Zhang},\ and\
  \citenamefont {Liao}}]{Bloczynski:2012en}%
  \BibitemOpen
  \bibfield  {author} {\bibinfo {author} {\bibfnamefont {J.}~\bibnamefont
  {Bloczynski}}, \bibinfo {author} {\bibfnamefont {X.-G.}\ \bibnamefont
  {Huang}}, \bibinfo {author} {\bibfnamefont {X.}~\bibnamefont {Zhang}}, \ and\
  \bibinfo {author} {\bibfnamefont {J.}~\bibnamefont {Liao}},\ }\href {\doibase
  10.1016/j.physletb.2012.12.030} {\bibfield  {journal} {\bibinfo  {journal}
  {Phys. Lett.}\ }\textbf {\bibinfo {volume} {B718}},\ \bibinfo {pages} {1529}
  (\bibinfo {year} {2013})}\BibitemShut {NoStop}%
\end{thebibliography}%

\end{document}